\documentclass[preprint,superscriptaddress]{revtex4-1}
\usepackage{graphicx}
\usepackage{bm,bbm}
\usepackage{amsmath,amsfonts,amssymb}
\usepackage[utf8]{inputenc}
\newcommand {\be} {\begin{equation}}
\newcommand {\ba} {\begin{eqnarray}}
\newcommand {\ee} {\end{equation}}
\newcommand {\ea} {\end{eqnarray}}

\begin{document}

\title{Gravitational form factor of the kaon in holographic QCD}

\author{Zhibo Liu}
\email{liu\_zhibo@hken.phys.nagoya-u.ac.jp}
\affiliation{School of Mathematics and Physics, University of South China, Hengyang 421001, People's Republic of China}
\affiliation{Department of Physics, Nagoya University, Nagoya 464-8602, Japan}
\author{Akira Watanabe}
\email{watanabe.akira@oshima-k.ac.jp}
\affiliation{School of Mathematics and Physics, University of South China, Hengyang 421001, People's Republic of China}
\affiliation{National Institute of Technology, Oshima College, Oshima 742-2193, Japan}

\date{\today}

\begin{abstract}
The gravitational form factor (GFF) of the kaon is investigated in a bottom-up holographic QCD model, in which the strange quark mass breaks the SU(3) flavor symmetry.
The probe energy ($Q^2$) dependence of the kaon GFF is explicitly shown and compared to that of the pion.
It is presented that our result shows the $1/Q^2$ behavior in the high energy region, which is consistent with the perturbative QCD prediction.
The gravitational radius of the kaon is also calculated, and it is found that the result is quite close to that of the pion but slightly smaller.
\end{abstract}

\maketitle

\section{Introduction}
\label{sec:introduction}
Perturbative QCD is a remarkably successful theory for studying the strong interaction in the high energy region, where the coupling is quite weak~\cite{Gross:1973id, Politzer:1973fx}.
However, effective approaches are essential to investigate physical quantities in the nonperturbative kinematic regions, such as hadron form factors, scattering cross sections, parton distribution functions and so on.
The hadronic matrix element of the energy-momentum tensor, known as the gravitational form factor (GFF), provides important information about the internal structure of hadrons and the dynamics of the gluon interaction.
GFFs have become a topic of great interest in the hadron physics field in recent years~\cite{Polyakov:2002yz, Belitsky:2002jp, Abidin:2008hn, Broniowski:2008hx, Abidin:2008ku, Abidin:2009aj, Abidin:2009hr, Ballon-Bayona:2017bwk, Kumano:2017lhr, Polyakov:2018exb, Freese:2019bhb, Polyakov:2019lbq, Krutov:2020ewr, Shi:2020pqe, Kim:2020lrs, Metz:2021lqv, Tong:2021ctu, Mamo:2022eui, Krutov:2022zgg, Hackett:2023rif, Guo:2023qgu, Hackett:2023nkr, Guo:2023pqw, Hatta:2023fqc, Xu:2023izo, Fujii:2024rqd}.
Various approaches have been developed and applied to calculate gravitational form factors so far.

The holographic QCD approach~\cite{Aharony:1999ti, Kruczenski:2003uq, Sakai:2004cn, Erlich:2005qh, deTeramond:2005su, Sakai:2005yt, Brodsky:2006uqa} is a nonperturbative method for studying QCD. It is based on the anti-de Sitter/conformal field theory (AdS/CFT) correspondence~\cite{Veneziano:1968yb, Maldacena:1997re, Gubser:1998bc, Witten:1998qj}, which is a duality between a gravitational theory in the AdS space and a CFT living on its boundary.
In order to describe QCD, which is not conformally invariant, additional ingredients are needed to break the conformal symmetry~\cite{Callan:1977gz}.
One way to achieve this is by introducing energy scale-dependent operators in the holographic dual theory.
In the bottom-up holographic QCD approach, energy scale-dependent operators are introduced to break the conformal invariance and capture the effects of the strong interactions at different energy scales.
These operators allow the holographic dual theory to reproduce the running of the QCD coupling constant and other non-conformal features of QCD.

Applying holographic QCD, researchers have been able to study various aspects of QCD, such as the spectrum and structure of hadrons~\cite{deTeramond:2005su, Erlich:2005qh, Brodsky:2006uqa, Brodsky:2007hb, Kwee:2007nq, Branz:2010ub, Gutsche:2011vb, Li:2013oda, Brodsky:2014yha, Gutsche:2017lyu, Ballon-Bayona:2017bwk, Lyubovitskij:2020gjz, Chen:2021wzj, Zhang:2021itx, Chen:2022goa}, properties of the quark-gluon plasma at high temperatures~\cite{Gursoy:2010fj, Bergman:2007wp}, the hadron form factors~\cite{ Kwee:2007dd, Grigoryan:2007wn, Grigoryan:2007vg, Abidin:2008hn, Abidin:2008ku, Abidin:2009aj, Abidin:2009hr, Ballon-Bayona:2017bwk, Abidin:2019xwu, Ahmed:2023zkk}, and the high energy scattering processes~\cite{ Polchinski:2001tt, Polchinski:2002jw, Brower:2006ea, Hatta:2007he, Pire:2008zf, Domokos:2009hm, Domokos:2010ma, Marquet:2010sf, Watanabe:2012uc, Watanabe:2013spa, Watanabe:2015mia, Anderson:2016zon, Watanabe:2018owy, Burikham:2019zbo, Watanabe:2019zny,  Xie:2019soz, Liu:2022out, Liu:2022zsa, Liu:2023tjr, Watanabe:2023rgp, Zhang:2023nsk, Watanabe:2023kki, Zhang:2024psj}.
It has been served as a powerful tool for understanding nonperturbative phenomena in QCD and has led to many interesting insights and developments in the field.
Especially for the GFFs, the results for the proton and pion obtained by the authors of Refs.~\cite{Abidin:2009hr, Abidin:2008hn} are consistent with the lattice QCD calculations~\cite{Hackett:2023rif, Hackett:2023nkr}, which strongly supports applications of the bottom-up holographic QCD models to analysis on the GFFs.

Then, it is natural to have interest in the analysis on the kaon structure.
The kaon was discovered in 1947 as the first strange particle, and has been regarded as one of the most important hadrons to understand various properties of the strong interaction in hadron physics.
The kaon has two important features, the Nambu-Goldstone boson nature and the fact that it includes a strange quark which explicitly breaks the SU(3) flavor symmetry.
From the former, it is naturally expected that the kaon may have the similar structure to that of the pion.
The explicit comparison between the resulting kaon GFF and pion's is presented in this paper, from which one can see the the size of the SU(3) flavor symmetry breaking effect.
It is expected that our understandings about the pion and kaon structure~\cite{Aguilar:2019teb, Roberts:2021nhw} will be further deepened via the experiments at the future Electron-Ion Colliders (EICs)~\cite{Accardi:2012qut, Aschenauer:2017jsk, Chen:2020ijn, AbdulKhalek:2021gbh, Anderle:2021wcy}.

The electromagnetic form factor of the kaon was studied in the bottom-up holographic QCD model by the authors of Ref.~\cite{Abidin:2019xwu}.
We extend it to the analysis on the GFF of the kaon in the present work.
The original bottom-up AdS/QCD model was extended to $\mathrm{SU(3)}_L \times \mathrm{SU(3)}_R$ with the broken flavor symmetry~\cite{Abidin:2009aj}.
We begin with the brief introduction of its formalism by explaining the connection between 4D operators $\mathcal{O}(x)$ and fields in the 5D bulk $\phi(x, z)$.
Subsequently, we present our analysis on the kaon GFF.
Our model involves three adjustable parameters, but we can utilize the values determined in Ref.~\cite{Abidin:2009aj} by fitting the hadron masses and decay constants.
It is shown that the probe energy ($Q^2$) dependence of the resulting kaon GFF is similar to that of the pion but some difference can be seen.
Our result shows the $1/Q^2$ behavior in the high energy region, which is consistent with the perturbative QCD prediction.
Using the resulting kaon GFF, the gravitational radius is also calculated, and it is found that the result is quite close to that of the pion but slightly smaller.

This paper is organized as follows.
In the next section, we briefly introduce the framework for the $N_f = 3$ case in holographic QCD and relevant meson pieces which encompass the masses, decay constants, and bulk profiles.
The three point function and extraction of the kaon GFF are presented with the numerical result in Sec.~\ref{sec:3}.
In Sec.~\ref{sec:4}, we show our calculation of the gravitational radius of the kaon, and then the conclusion of this work is given in Sec.~\ref{sec:conclusion}.

\section{Bottom-up holographic QCD model for $N_f = 3$}
\label{sec:2}
In the standard bottom-up holographic QCD prescription, the following operator/field correspondences are taken into account:
\begin{equation}
{J_V^a}_\mu(x) = \bar{q}(x)\gamma_\mu t^a q(x) \Leftrightarrow V_\mu^a(x,z), \nonumber
\end{equation}
\begin{equation}
{J_A^a}_\mu(x) = \bar{q}(x)\gamma_\mu \gamma_5t^a q(x) \Leftrightarrow A_\mu^a(x,z),\nonumber
\end{equation} 
\begin{equation}
T_{\mu\nu}(x) \Leftrightarrow h_{\mu\nu}(x,z),
\end{equation} 
Here $h_{\mu\nu}(x,z)$ denotes the graviton (metric fluctuation) in the AdS$_5$ bulk, which is holographically dual to the four-dimensional energy-momentum tensor $T_{\mu\nu}(x)$. Its boundary value $h^{\mu\nu}_0(x)$ acts as the source for $T_{\mu\nu}(x)$ in the generating functional.
where $t^a=\sigma^a/2$ are the SU(3) generators, $\sigma^a$ are Gell-Mann matrices, and for $h_{\mu\nu}$ the transverse-traceless gauge is applied.
In this study, we adopt the hard-wall model, in which a sharp cutoff is imposed in the infrared (IR) region to introduce the QCD scale.
The background AdS space metric is
\begin{equation}
ds^2 = \frac{1}{z^2}\left( \eta_{\mu\nu}dx^\mu dx^\nu - dz^2 \right),
\end{equation}
where $\eta_{\mu\nu} = \mathrm{diag}(1,-1,-1,-1)$, $\epsilon < z < z_0$, and $z_0 \sim 1/\Lambda_{\mathrm{QCD}}$ is the cutoff.

The action in the 5D AdS space is~\cite{Erlich:2005qh}
\begin{align}
\label{action}
S_{5D} =-\int_{\epsilon}^{z0} d^5x \sqrt{-g}\bigg\{&\mathrm{Tr}\left[|DX|^2-3|X|^2 -\frac{1}{4g_5^2}(F_L^2+F_R^2)  \right]  \bigg\},
\end{align}
where $g$ = det$[g_{MN}]$ and the complex scalar field $X$ describes the chiral symmetry breaking, which corresponds to the 4D operator $\bar{q}_Rq_L$,
\begin{equation}
X(x,z) = X_0(z)\mathrm{exp}(2it^a\pi^a).
\end{equation}
The covariant derivative of this field is expressed as
\begin{equation}
D^M X = \partial^M X + iXA_R^M - iXA_L^M X.
\end{equation}
The solution to the equation of motion for the bulk scalar is given by
\begin{equation}
X_0 = a_1 \zeta z + a_3 z^3/\zeta,
\end{equation}
where the first term on the right-hand side represents the chiral symmetry breaking caused by the presence of non-zero quark mass, the second term describes the quark condensate, and $\zeta = \sqrt{N_c}/2\pi$ is the normalization factor.
$a_1$ and $a_3$ are expressed as
\begin{equation}
a_1 = \frac{1}{2}M \mathbbm{1} =\frac{1}{2} \left(
\begin{matrix}
m_q~&   & \\
   &m_q~& \\
   &   &m_s
\end{matrix}
\right),
\end{equation}
\begin{equation}
a_3 = \frac{1}{2}\Sigma\mathbbm{1}=\frac{1}{2}  \left(
\begin{matrix}
\sigma_q~&           & \\
        &\sigma_q  ~ & \\
        &           &\sigma_s
\end{matrix}
\right),
\end{equation}
where the SU(2) isospin symmetry is assumed.
Then, we define that
\begin{equation}
v_q = m_q\zeta z + \frac{\sigma}{\zeta} z^3~~~\mathrm{and}~~~v_s = m_s\zeta z + \frac{\sigma}{\zeta} z^3,
\end{equation}
in which $\sigma_q = \sigma_s = \sigma$ is assumed. The parameter $z_0 = (322.5~\mathrm{MeV})^{-1}$ is determined by fitting the $\rho$ meson mass~\cite{Erlich:2005qh}.

The field strength tensors $F_{MN}^{L}$ and $F_{MN}^{R}$ are expressed as
\begin{equation}
F_{MN}^{L} = \partial_M L_N - \partial_N L_M - i[L_M,L_N],
\end{equation}
\begin{equation}
F_{MN}^{R} = \partial_M R_N - \partial_N R_M - i[R_M,R_N],
\end{equation}
where $L^M = L^{Ma}t^a$. The vector and axial-vector fields can be expressed with $L_M$ and $R_M$:
\begin{equation}
V_M = (L_M + R_M)/2,
\end{equation}
\begin{equation}
A_M = (L_M - R_M)/2.
\end{equation}
Taking the fields from the action up to second order,
\begin{align}
\label{2_action}
S^{(2)} = \int d^5x  \Bigg[ &\sum_{a=1}^{8} \frac{-1}{4g_5^2z}(\partial_M V_N^a - \partial_N V_M^a)^2 + \frac{(M_V^a)^2}{2z^3} (V_M^a)^2    \nonumber
\\& -\frac{1}{4g_5^2 z} (\partial_M A_N^a - \partial_N A_M^a)^2  + \frac{(M_A^a)^2}{2z^3}(\partial_M \pi^a - A_M^a)^2\Bigg],
\end{align}
where the mass terms are defined by
\begin{equation}
(M_V^a)^2 \delta^{ab} = 2\mathrm{Tr}\big( [t^a,X_0][ t^b,X_0 ] \big),
\end{equation}
\begin{equation}
(M_A^a)^2 \delta^{ab} = 2\mathrm{Tr}\big( \{t^a,X_0\}\{ t^b,X_0 \} \big).
\end{equation}
Expanding the above equations yields
\begin{equation}
(M_V^a)^2 = 
\left\{ \begin{matrix}
0&~~~~~ a=1,2,3 ,\\
\frac{1}{4}(v_s-v_q)^2& ~~~~~~~~a=4,5,6,7, \\
0 & a=8,\\
\end{matrix} \right.
 \end{equation}
\begin{equation}
(M_A^a)^2 = 
\left\{ \begin{matrix}
v_q^2&~~~~~ a=1,2,3, \\
\frac{1}{4}(v_s+v_q)^2& ~~~~~~~~a=4,5,6,7, \\
\frac{1}{3}(v_q^2 + 2v_s^2) & a=8.\\
\end{matrix} \right.
 \end{equation}

Adopting the axial-like gauge, $A_z =0$ and $V_z =0$, and taking variations for $A_M^a(x,z)$ and $V_M^a(x,z)$ respectively, one obtains the equations of motion and the Fourier transform, which are expressed as
\begin{equation}
\label{EM_v}
\partial_z\left( \frac{1}{z}\partial_zV_{\mu\perp}^a(q,z)\right) + \frac{q^2-\alpha^a}{z}V_{\mu\perp}^a(q,z) =0,
\end{equation}
\begin{equation}
\label{EM_A}
z\partial_z\left(\frac{1}{z}\partial_z {A_\mu^a}_\perp(q,z)\right) - \beta^a(z){A_\mu^a}_\perp(q,z) = q^2 {A_\mu^a}_\perp(q,z),
\end{equation}
\begin{equation}
\label{EM_phi}
\partial_z \left( \frac{1}{z}\partial_z \phi^a(q,z) \right) =\frac{\beta^a(z)}{z}\big( \phi^a(q,z) - \pi^a(q,z) \big),
\end{equation}
\begin{equation}
\label{EM_pi}
\beta^a(z)\partial_z\pi^a(q,z)=q^2\partial_z\phi^a(q,z),
\end{equation}
where 
\begin{equation}
\alpha^a(z) = \frac{g_5^2 (M_V^a)^2}{z^2}, \ \ \ \beta^a(z) = \frac{g_5^2 (M_A^a)^2}{z^2}.
\end{equation}
Here $V_{\mu\perp}^a(q,z)$ are the 4D Fourier transform of $V_{\mu\perp}^a(x,z)$, $V_{\mu\perp}^a(q,z) = \int d^4q e^{iqx}V_{\mu\perp}^a(x,z)$,
and the boundary conditions are $V(q,\varepsilon)= 1$, $\partial_zV^a(q,z_0)=0$,
$\phi^a(q,\varepsilon)=0$, $\pi^a(q,\varepsilon)=-1$, and $\partial_z\phi^a(q,z_0)=0$.
The fields $A_\mu(q,z)$ can be decomposed  into the transverse part $A_{\mu\perp}$ and the longitudinal part $\partial_\mu\phi$,
\begin{equation}
A_\mu(q,z) = A_{\mu\perp}(q,z)+\partial_\mu\phi(q,z).
\end{equation}
Similarly to the vector fields, $A_{\mu\perp}^a(q,z)$, $\pi^a(q,z)$, and $\phi^a(q,z)$ are the 4D Fourier transform of $A_{\mu\perp}^a(x,z)$, $\pi^a(x,z)$, and $\phi^a(x,z)$, respectively.

The fields $V_\mu^a(q,z)$, $A_\mu^a(q,z)$, and $\pi^a(q,z)$ can be expressed as
\begin{equation}
V_{\perp \mu}^a(q,z) = V(q,z) V_\mu^{a0}(q),\nonumber
\end{equation}
\begin{equation}
A_\mu^a(q,z) = A_{\mu\perp}^{a0}(q)\mathcal{A}(q,z) + \frac{p^\alpha p_\mu}{p^2}A_{\alpha\Vert }^{a0}(q) \phi(q,z),\nonumber
\end{equation}
\begin{equation}
\pi^a(q,z) = \frac{1}{iq_\mu}A_{\mu\Vert}^{a0}(q)\pi(q,z),
\end{equation}
where $V_\mu^{a0}(q)$ corresponds to the Fourier transform of the source function of the 4D vector current operator $J_{V\mu}^a(x)$, and $A_{\mu\perp}^{a0}(q)$ and $A_{\mu\Vert}^0(q)$ are the Fourier transform of the source functions of the 4D axial current operators $J_{A,\mu \perp}^a(x)$ and $J_{A,\mu \Vert}^a(x)$, respectively.

According to the AdS/QCD correspondence, the two point functions can be calculated  by taking the derivative over $V_\mu^{a0}(q)$ twice for the action as
\begin{equation}
\label{TPF}
\langle 0|\mathcal{T} J_{V\perp}^{a\mu}(p_1)  J_{V\perp}^{b\nu}(p_2) |0\rangle = -i \frac{\delta^2 S_{VV}}{\delta V_{\mu\perp}^{a0}(p_1) \delta V_{\nu\perp}^{b0}(p_2)},
\end{equation}
where 
\begin{equation}
\label{TPA}
S_{VV} = -\int \frac{d^4q}{(2\pi^2)} V^{0\mu}(q)V^0_\mu(q) \frac{\partial_z V(q,z)}{2g_5^2z}\bigg|_{z=\epsilon}
\end{equation}
is obtained from Eq.~\eqref{2_action}.
The definition of the decay constant of vector meson $f_V$ is written as
\begin{equation}
\label{definition_decay}
\langle J_{V\perp}^{\mu a}| V^{b}(p,\epsilon) \rangle = f_{V}\epsilon^\mu \delta^{ab},
\end{equation} 
where $\epsilon$ is the polarization vector.

\begin{figure}[tb]
\centering
\includegraphics[width=0.6\textwidth]{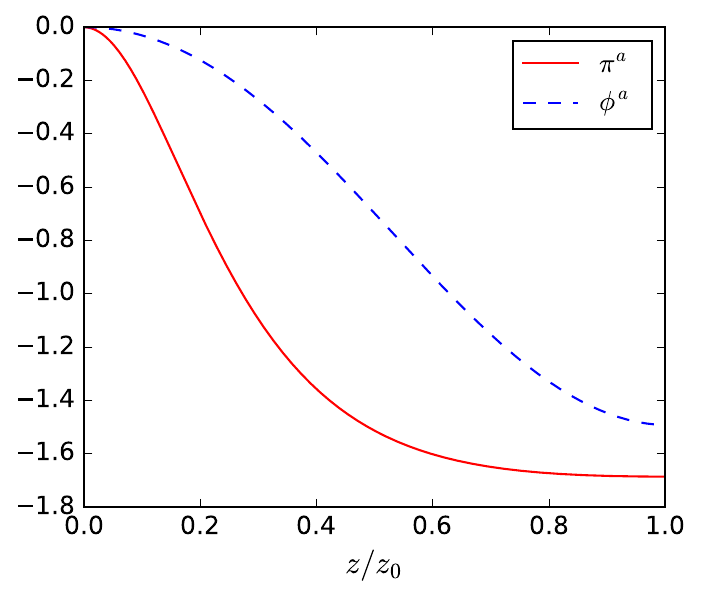}
\caption{
The kaon bulk profiles $\pi^a$ and $\phi^a$ as a function of $z$, which correspond to $M_K = 495.7$~MeV for a = 4,5,6,7.
}
\label{WF_K}
\end{figure}

The bulk-to-boundary propagator $V^a$, the solution to Eq.~\eqref{EM_v}, can be expressed using the Green’s function formalism as
\begin{equation}
\label{green function}
V^a(q,z) = \sum_n \frac{-g_5f_{V,n}\psi_{V,n}^a(z)}{q^2 - m_{V,n}^2},
\end{equation}
where $\psi_{V,n}^a(z)$ is the bulk profile dual to the $n$-th vector meson, the bulk profile also satisfies the equation of motion, Eq.~\eqref{EM_v}, and the boundary conditions are the same as those for $V_{\mu\perp}(q,z)$. 
Their eigenvalues $q^2 = (m_n^a)^2$ are the squared mass of the corresponding mesons.
The normalization condition for $\psi_{V,n}^a(z)$ is $\int (dz/z) { (\psi_{V,n}^a (z))^2} =1$. Combining Eqs.~\eqref{TPF}, \eqref{TPA}, \eqref{definition_decay}, and \eqref{green function}, the decay constant of the $n$-th vector meson is written as
\begin{equation}
\label{decay_v}
f_{Vn} = \frac{\partial_z \psi_{V,n}(z)}{g_5 z}\bigg |_{z = \epsilon}.
\end{equation}
Similarly, the bulk-to-boundary propagators, $\mathcal{A}^a(q,z)$, $\pi^a(q,z)$, and $\phi^a(q,z)$, can be represented as the sum over the axial vector meson and pseudoscalar meson poles,
\begin{equation}
\mathcal{A}^a(q,z) = \sum_n \frac{ -g_5 f_{A,n}^a \psi_{A,n}(z) }{ q^2 - (m_{A,n}^a)^2 }, \nonumber
\end{equation}
\begin{equation}
\phi^a(q,z) = \sum_n \frac{ -g_5 (m_{P,n}^a)^2 f_{P,n}^a \phi_{n}(z) }{ q^2 - (m_{P,n}^a)^2 }, \nonumber
\end{equation}
\begin{equation}
\pi^a(q,z) = \sum_n \frac{ -g_5 (m_{P,n}^a)^2 f_{P,n}^a \pi_{n}(z) }{ q^2 - (m_{P,n}^a)^2 }.
\end{equation}
The decay constants of the axial vector meson and pseudoscalar meson are expressed as
\begin{equation}
f_{A,n} = \frac{\partial_z \psi_{A,n}(z)}{g_5 z}\bigg|_{z=\epsilon},
\end{equation}
\begin{equation}
\label{decay_pi}
f_{P,n} = -\frac{\partial_z \phi_{n}(z)}{g_5 z}\bigg|_{z=\epsilon},
\end{equation}
where $\psi_{A,n}$ and $\phi_{n}$ are bulk profiles of the axial vector meson and pseudoscalar meson, respectively.
Moreover, the normalization condition for $\phi_n(z)$ is \cite{Abidin:2009aj}
\begin{equation}
\int dz \frac{z}{\beta^a(z)} (y_n^a (z))^2 = \frac{1}{(m_n^a)^2},
\end{equation}
where $y^a_n(z) = \partial_z\phi_n^a/z$.
Using Eqs.~\eqref{EM_v}, \eqref{EM_phi}, \eqref{EM_pi}, \eqref{decay_v}, \eqref{decay_pi} and Gell-Mann--Oakes--Renner relation $m_\pi^2f_\pi^2 = 2m_q\sigma_q$, the authors of Ref.~\cite{Abidin:2009aj} determined the parameters $m_q = 8.31$ MeV and $\sigma_q = (213.7~\mathrm{MeV})^3$ by fitting to the pion mass $M_\pi = 139.6$ MeV and decay constant $f_\pi$ = 92.4 MeV. The strange quark mass $m_s$ = 188.5 MeV and the kaon decay constant $f_K$ = 104~MeV are determined by fitting to the kaon mass $M_K$ = 495.7 MeV.
With these parameter values, the kaon bulk profiles $\pi^a(z)$ and $\phi^a(z)$ are obtained, which are shown in Fig.~\ref{WF_K}.
%

\section{Gravitational form factor of the kaon}
\label{sec:3}
Our calculation of the kaon GFF is presented in this section.
The stress tensor matrix element of pseudoscalar $\langle \pi_n^a(x)| T_{\mu\nu}(y)|\pi_n^b(w) \rangle$ can be extracted from the three point function
\begin{equation}
\langle 0 | \mathcal{T} J_{A\Vert}^{a\alpha}(x) \hat{T}_{\mu\nu}(y)J_{A\Vert}^{b\beta}(w)|0 \rangle = -\frac{2\delta^3 S_{A_\Vert G A_\Vert}}{\delta A_{\alpha\Vert}^{a0}(x) \delta h^{\mu\nu 0 }(y) \delta A_{\beta\Vert}^{b0}(w)  }.
\end{equation}
The relevant part in the action, Eq.~\eqref{action}, is written as
\begin{align}
S_{A_\Vert G A_\Vert } = \int & d^5x  \bigg[ -\frac{v^a(z)^2 h^{\mu\nu}}{2z^3}(\partial_\mu\pi^a - A_\mu^a)(\partial_\nu\pi^a - A_\nu^a)   - \frac{1}{2g_5^2z}h^{\mu\nu}F_{\nu z}^aF_{\mu z}^a    \bigg],
\end{align}
where 
\begin{equation}
v^a = \left\{
\begin{matrix}
v_q&   a= 1,2,3, \\
 \frac{1}{2}(v_q + v_s)  & ~~~a = 4,5,6,7. \\
\end{matrix}
\right.
\end{equation}
Using this representation and considering the Fourier transform,
\begin{align}
S_{A_\Vert G A_\Vert } = &\int \frac{d^4 u d^4 v d^4 w}{(2\pi)^{4\times 3}} dz (2\pi)^4 \delta^{(4)}(u+v+w) \Bigg\{ -\frac{(v^a(z))^2 h^{\mu\nu0} \mathcal{H}(w,z)}{ 2z^3 } \nonumber
\\ & \times \bigg( \frac{ u^\alpha u_\mu }{u^2} A_{\alpha\Vert}^0(u)\pi(q,z) - \frac{ u^\alpha u_\mu }{u^2} A_{\alpha\Vert}^0(u) \phi(q,z) \bigg) 
\\ & \times \bigg( \frac{ u^\beta u_\mu }{u^2} A_{\beta\Vert}^0(u)\pi(q,z) - \frac{ u^\beta u_\mu }{u^2} A_{\beta\Vert}^0(u) \phi(q,z) \bigg) \nonumber
\\ & -\frac{1}{2g_5^2z} h^{\mu\nu0}(w)  \mathcal{H}(w,z) \partial_z\bigg[ \frac{v^\beta v_\mu}{v^2} A_{\beta\Vert}^0(v) \phi(q,z) \bigg] \partial_z\bigg[ \frac{v^\beta v_\nu}{v^2} A_{\beta\Vert}^0(v) \phi(q,z) \bigg]  \Bigg\}.
\end{align}
The three point function can be extracted from this equation, which is given by
\begin{align}
\label{vertex}
&~~~~\langle 0 | \mathcal{T}J^\alpha(p_2) T_{\mu\nu}(q) J^{\beta}(p_1)|0\rangle \nonumber
\\& =  \frac{-2\delta^3 S_{A_\Vert G A_\Vert}}{\delta A_{\alpha\Vert}^{a0}(p_2) \delta h^{\mu\nu 0 }(q) \delta A_{\beta\Vert}^{b0}(p_1)}  \nonumber
 \\&= (2\pi)^4 \delta^{(4)}(p_1+q-p_2) \frac{p_2^{\alpha} p_{2\mu}}{p_2^2 - m^2_K}\times \frac{p_1^{\beta} p_{1\nu}}{p_1^2 - m^2_K} \nonumber
\\ & ~~~~\times 2\int dz \frac{\mathcal{H}(q,z)}{2z^3}\Bigg[ v^a(z)\bigg( \pi(z)-\phi(z) \bigg)^2  + \frac{1}{2g_5^2z} \bigg(\partial_z [\phi(z)]\bigg)^2 \Bigg],
\end{align}
and the matrix element is obtained as
\begin{align}
&\langle K(p_2)|T_{\mu\nu}(0)|K(p_1)\rangle \nonumber
\\ =& \lim\limits_{p_1^2\rightarrow m_K^2 \atop p_2^2\rightarrow m_K^2}
\frac{(p_1^2-m_K^2)({p_2}^2-m_K^2)}{p^\alpha p^\beta f_K^2}\times \langle 0 | \mathcal{T}J^\alpha(-p_2) T_{\mu\nu}(q) J^{\beta}(p_1)|0\rangle \nonumber
\\=& p_\mu p_\nu A_K(Q^2).
\end{align}
The GFF of the kaon is extracted from this matrix element and can be expressed as
\begin{align}
A_{K}(Q^2) =\int& dz \mathcal{H}(Q,z)\Bigg[ \frac{g_5^2 (v^a (z))^2}{z^3}\bigg( \pi(z)-\phi(z) \bigg)^2  + \frac{1}{z} \bigg(\partial_z [\phi(z)]\bigg)^2 \Bigg],
\end{align}
where $\mathcal{H}(Q,z)$ is the bulk-to-boundary propagator of the graviton.

\begin{figure}[tb]
\centering
\includegraphics[width=0.6\textwidth]{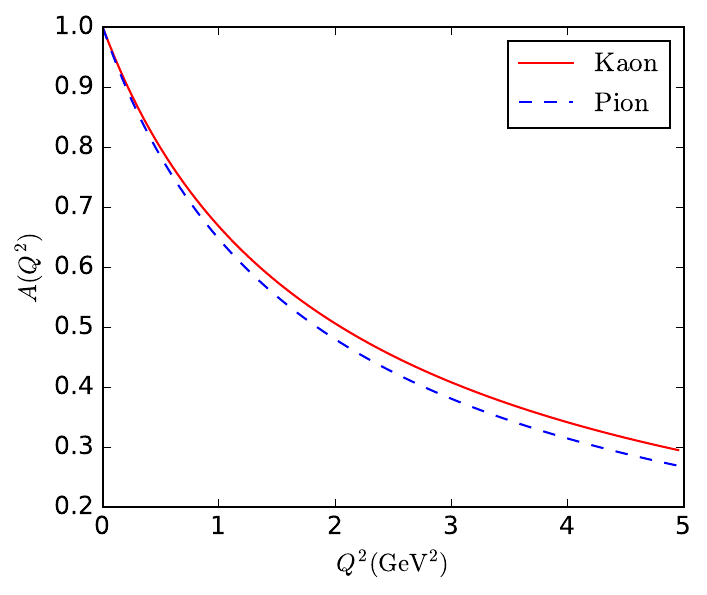}
\caption{
The GFF as a function of $Q^2$.
The solid and dashed curves represent our kaon result and the preceding study result for the pion~\cite{Abidin:2008hn}, respectively.
}
\label{GFF}
\end{figure}

The 5D metric is perturbed from the flat background,
\begin{equation}
ds^2 = \frac{1}{z^2}\big( (\eta_{\mu\nu} + h_{\mu\nu})dx^\mu dx^\nu - dz^2 \big),
\end{equation}
where $h_{\mu\nu}(q,z) = \mathcal{H}(q,z)h_{\mu\nu}^0(q)$, in which $h_{\mu\nu}^0(q)$ is the Fourier transform of the source function of the graviton.
Here we consider perturbations of the AdS$_5$ metric around the hard-wall background. In the transverse-traceless gauge,
$q^\mu h_{\mu\nu}(q,z) = 0$ and $h^\mu_{\;\mu}(q,z) = 0$, the graviton mode
$h_{\mu\nu}(q,z)$ satisfies the following wave equation:
\begin{equation}
  z^3 \partial_z\!\left(\frac{1}{z^3}\partial_z h_{\mu\nu}(q,z)\right)
  + q^2 h_{\mu\nu}(q,z) = 0,
  \label{eq:44}
\end{equation}
which is the standard equation of motion for a massless spin-2 field (graviton)
propagating in AdS$_5$ used in the hard-wall model.
For the spacelike momentum transfer $q^2 = -Q^2 < 0$, the solution to this equation is given by
\begin{equation}
\mathcal{H}(Q,z) = \frac{1}{2} Q^2z^2 \bigg( \frac{K_1(Qz_0)}{I_1(Qz_0)}I_2(Qz) + K_2(Qz) \bigg).
\end{equation}
The $Q^2$ dependence of $A_K$ is shown in Fig.~\ref{GFF} together with the pion GFF $A_\pi$ which is taken from Ref.~\cite{Abidin:2008hn}.
It can be seen from the figure that the behaviors of the both GFFs are similar to each other, but $A_K$ decreases slightly slower than $A_\pi$ with $Q^2$.

In the $Q^2 \rightarrow \infty$ limit, the kaon GFF can be expressed as
\begin{equation}
A_K(Q^2) = \frac{4\rho(\epsilon)}{Q^2} = \frac{16\pi^2 f_K^2}{Q^2}.
\end{equation}
Hence, $Q^2A_K$ is found to be constant:
\begin{equation}
Q^2 A_K(Q^2) = 16\pi^2 f_K^2 = 1.71~\mathrm{GeV^2}.
\end{equation}
This $1/Q^2$ behavior in the high energy region is consistent with the perturbative QCD prediction~\cite{Lepage:1980fj, Tanaka:2018wea, Tong:2021ctu, Tong:2022zax}.

\section{Gravitational radius of the kaon}
\label{sec:4}
In this section, we present our calculation of the gravitational radius of the kaon.
The root-mean square radius is defined as
\begin{equation}
\langle r_K^2\rangle _{\mathrm{grav}} = -6\frac{\partial A_K(Q^2)}{\partial Q^2}\bigg|_{Q^2=0}.
\end{equation}
Expanding $\mathcal{H}(Q,z)$ at $Qz_0 \rightarrow 0$, this equation can be rewritten in the $Q^2 \rightarrow 0$ limit,
\begin{equation}
\langle r_K^2\rangle _{\mathrm{grav}} = \frac{6}{4}\int dz z^3 \bigg( 1- \frac{z^2}{2z_0^2} \bigg)\rho(z),
\end{equation}
where 
\begin{equation}
\rho(z) = \frac{(v^a (z))^2}{z^3}\big( \pi-\phi \big)^2 + \frac{1}{g_5^2z} \big(\partial_z \phi\big)^2.
\end{equation}
Then, the gravitational radius can be calculated as 
\begin{equation}
\langle r_K^2\rangle _{\mathrm{grav}} = 0.12~\mathrm{(fm)}^2 = (0.35~\mathrm{fm})^2.
\end{equation}
Our result is substantially smaller than the charge radius of the kaon $\langle r_K^2 \rangle _{E} = (0.56 \pm 0.031~\mathrm{fm})^2$~\cite{ParticleDataGroup:2022pth}, and slightly smaller than the preceding study result for the pion $\langle r_\pi^2\rangle _{\mathrm{grav}} = (0.36~\mathrm{fm})^2$~\cite{Abidin:2008hn}, although this can be seen from Fig.~\ref{GFF} also.
Therefore, it is found that the observed size of the SU(3) flavor symmetry breaking effect in the present bottom-up holographic QCD model is quite small despite the mass differences between the up/down and strange quarks and between the pion and kaon.

\section{Conclusion}
\label{sec:conclusion}
We have investigated the kaon GFF in the bottom-up holographic QCD model with a sharp cutoff in the IR region for the background AdS space.
Since the strange quark mass breaks the SU(3) flavor symmetry, the size of this symmetry breaking effect can be seen via the comparison between the resulting kaon GFF and the preceding study result for the pion.
Beginning with the brief introduction of the fundamentals of the framework, we have explicitly demonstrated how one can obtain the GFF from the model action.

We have presented the energy dependence of the resulting kaon GFF, which is found to be similar to that of the pion, but slightly weaker.
Our calculation shows the $1/Q^2$ behavior in the high energy region, which is consistent with the perturbative QCD prediction.
The gravitational radius of the kaon has also been calculated, and we have obtained $\langle r_K^2\rangle _{\mathrm{grav}} = (0.35~\mathrm{fm})^2$, which is close to the result for the pion, but slightly smaller.
Through the study, it is found that the size of the SU(3) flavor symmetry breaking effect in the present holographic QCD model is quite small despite the mass differences between the up/down and strange quarks and between the pion and kaon.
Since this is one of the most important and interesting problems in high energy physics, to pin down this, further studies are certainly needed.
Future experiments at the EICs will help enhance the discussions and deepen our understandings about the strong interaction.

\section*{Acknowledgments}
We would like to thank Yidian Chen, Hai-Yang Cheng, Masayasu Harada, and Hsiang-nan Li for fruitful discussions.
This work was supported in part by the National Natural Science Foundation of China under Grant No. 12375137.

\bibliography{references}

\begin{thebibliography}{92}%
\makeatletter
\providecommand \@ifxundefined [1]{%
 \@ifx{#1\undefined}
}%
\providecommand \@ifnum [1]{%
 \ifnum #1\expandafter \@firstoftwo
 \else \expandafter \@secondoftwo
 \fi
}%
\providecommand \@ifx [1]{%
 \ifx #1\expandafter \@firstoftwo
 \else \expandafter \@secondoftwo
 \fi
}%
\providecommand \natexlab [1]{#1}%
\providecommand \enquote  [1]{``#1''}%
\providecommand \bibnamefont  [1]{#1}%
\providecommand \bibfnamefont [1]{#1}%
\providecommand \citenamefont [1]{#1}%
\providecommand \href@noop [0]{\@secondoftwo}%
\providecommand \href [0]{\begingroup \@sanitize@url \@href}%
\providecommand \@href[1]{\@@startlink{#1}\@@href}%
\providecommand \@@href[1]{\endgroup#1\@@endlink}%
\providecommand \@sanitize@url [0]{\catcode `\\12\catcode `\$12\catcode
  `\&12\catcode `\#12\catcode `\^12\catcode `\_12\catcode `\%12\relax}%
\providecommand \@@startlink[1]{}%
\providecommand \@@endlink[0]{}%
\providecommand \url  [0]{\begingroup\@sanitize@url \@url }%
\providecommand \@url [1]{\endgroup\@href {#1}{\urlprefix }}%
\providecommand \urlprefix  [0]{URL }%
\providecommand \Eprint [0]{\href }%
\providecommand \doibase [0]{http://dx.doi.org/}%
\providecommand \selectlanguage [0]{\@gobble}%
\providecommand \bibinfo  [0]{\@secondoftwo}%
\providecommand \bibfield  [0]{\@secondoftwo}%
\providecommand \translation [1]{[#1]}%
\providecommand \BibitemOpen [0]{}%
\providecommand \bibitemStop [0]{}%
\providecommand \bibitemNoStop [0]{.\EOS\space}%
\providecommand \EOS [0]{\spacefactor3000\relax}%
\providecommand \BibitemShut  [1]{\csname bibitem#1\endcsname}%
\let\auto@bib@innerbib\@empty
\bibitem [{\citenamefont {Gross}\ and\ \citenamefont
  {Wilczek}(1973)}]{Gross:1973id}%
  \BibitemOpen
  \bibfield  {author} {\bibinfo {author} {\bibfnamefont {D.~J.}\ \bibnamefont
  {Gross}}\ and\ \bibinfo {author} {\bibfnamefont {F.}~\bibnamefont
  {Wilczek}},\ }\href {\doibase 10.1103/PhysRevLett.30.1343} {\bibfield
  {journal} {\bibinfo  {journal} {Phys. Rev. Lett.}\ }\textbf {\bibinfo
  {volume} {30}},\ \bibinfo {pages} {1343} (\bibinfo {year}
  {1973})}\BibitemShut {NoStop}%
\bibitem [{\citenamefont {Politzer}(1973)}]{Politzer:1973fx}%
  \BibitemOpen
  \bibfield  {author} {\bibinfo {author} {\bibfnamefont {H.~D.}\ \bibnamefont
  {Politzer}},\ }\href {\doibase 10.1103/PhysRevLett.30.1346} {\bibfield
  {journal} {\bibinfo  {journal} {Phys. Rev. Lett.}\ }\textbf {\bibinfo
  {volume} {30}},\ \bibinfo {pages} {1346} (\bibinfo {year}
  {1973})}\BibitemShut {NoStop}%
\bibitem [{\citenamefont {Polyakov}(2003)}]{Polyakov:2002yz}%
  \BibitemOpen
  \bibfield  {author} {\bibinfo {author} {\bibfnamefont {M.~V.}\ \bibnamefont
  {Polyakov}},\ }\href {\doibase 10.1016/S0370-2693(03)00036-4} {\bibfield
  {journal} {\bibinfo  {journal} {Phys. Lett. B}\ }\textbf {\bibinfo {volume}
  {555}},\ \bibinfo {pages} {57} (\bibinfo {year} {2003})},\ \Eprint
  {http://arxiv.org/abs/hep-ph/0210165} {arXiv:hep-ph/0210165} \BibitemShut
  {NoStop}%
\bibitem [{\citenamefont {Belitsky}\ and\ \citenamefont
  {Ji}(2002)}]{Belitsky:2002jp}%
  \BibitemOpen
  \bibfield  {author} {\bibinfo {author} {\bibfnamefont {A.~V.}\ \bibnamefont
  {Belitsky}}\ and\ \bibinfo {author} {\bibfnamefont {X.}~\bibnamefont {Ji}},\
  }\href {\doibase 10.1016/S0370-2693(02)02025-7} {\bibfield  {journal}
  {\bibinfo  {journal} {Phys. Lett. B}\ }\textbf {\bibinfo {volume} {538}},\
  \bibinfo {pages} {289} (\bibinfo {year} {2002})},\ \Eprint
  {http://arxiv.org/abs/hep-ph/0203276} {arXiv:hep-ph/0203276} \BibitemShut
  {NoStop}%
\bibitem [{\citenamefont {Abidin}\ and\ \citenamefont
  {Carlson}(2008{\natexlab{a}})}]{Abidin:2008hn}%
  \BibitemOpen
  \bibfield  {author} {\bibinfo {author} {\bibfnamefont {Z.}~\bibnamefont
  {Abidin}}\ and\ \bibinfo {author} {\bibfnamefont {C.~E.}\ \bibnamefont
  {Carlson}},\ }\href {\doibase 10.1103/PhysRevD.77.115021} {\bibfield
  {journal} {\bibinfo  {journal} {Phys. Rev. D}\ }\textbf {\bibinfo {volume}
  {77}},\ \bibinfo {pages} {115021} (\bibinfo {year} {2008}{\natexlab{a}})},\
  \Eprint {http://arxiv.org/abs/0804.0214} {arXiv:0804.0214 [hep-ph]}
  \BibitemShut {NoStop}%
\bibitem [{\citenamefont {Broniowski}\ and\ \citenamefont
  {Ruiz~Arriola}(2008)}]{Broniowski:2008hx}%
  \BibitemOpen
  \bibfield  {author} {\bibinfo {author} {\bibfnamefont {W.}~\bibnamefont
  {Broniowski}}\ and\ \bibinfo {author} {\bibfnamefont {E.}~\bibnamefont
  {Ruiz~Arriola}},\ }\href {\doibase 10.1103/PhysRevD.78.094011} {\bibfield
  {journal} {\bibinfo  {journal} {Phys. Rev. D}\ }\textbf {\bibinfo {volume}
  {78}},\ \bibinfo {pages} {094011} (\bibinfo {year} {2008})},\ \Eprint
  {http://arxiv.org/abs/0809.1744} {arXiv:0809.1744 [hep-ph]} \BibitemShut
  {NoStop}%
\bibitem [{\citenamefont {Abidin}\ and\ \citenamefont
  {Carlson}(2008{\natexlab{b}})}]{Abidin:2008ku}%
  \BibitemOpen
  \bibfield  {author} {\bibinfo {author} {\bibfnamefont {Z.}~\bibnamefont
  {Abidin}}\ and\ \bibinfo {author} {\bibfnamefont {C.~E.}\ \bibnamefont
  {Carlson}},\ }\href {\doibase 10.1103/PhysRevD.77.095007} {\bibfield
  {journal} {\bibinfo  {journal} {Phys. Rev. D}\ }\textbf {\bibinfo {volume}
  {77}},\ \bibinfo {pages} {095007} (\bibinfo {year} {2008}{\natexlab{b}})},\
  \Eprint {http://arxiv.org/abs/0801.3839} {arXiv:0801.3839 [hep-ph]}
  \BibitemShut {NoStop}%
\bibitem [{\citenamefont {Abidin}\ and\ \citenamefont
  {Carlson}(2009{\natexlab{a}})}]{Abidin:2009aj}%
  \BibitemOpen
  \bibfield  {author} {\bibinfo {author} {\bibfnamefont {Z.}~\bibnamefont
  {Abidin}}\ and\ \bibinfo {author} {\bibfnamefont {C.~E.}\ \bibnamefont
  {Carlson}},\ }\href {\doibase 10.1103/PhysRevD.80.115010} {\bibfield
  {journal} {\bibinfo  {journal} {Phys. Rev. D}\ }\textbf {\bibinfo {volume}
  {80}},\ \bibinfo {pages} {115010} (\bibinfo {year} {2009}{\natexlab{a}})},\
  \Eprint {http://arxiv.org/abs/0908.2452} {arXiv:0908.2452 [hep-ph]}
  \BibitemShut {NoStop}%
\bibitem [{\citenamefont {Abidin}\ and\ \citenamefont
  {Carlson}(2009{\natexlab{b}})}]{Abidin:2009hr}%
  \BibitemOpen
  \bibfield  {author} {\bibinfo {author} {\bibfnamefont {Z.}~\bibnamefont
  {Abidin}}\ and\ \bibinfo {author} {\bibfnamefont {C.~E.}\ \bibnamefont
  {Carlson}},\ }\href {\doibase 10.1103/PhysRevD.79.115003} {\bibfield
  {journal} {\bibinfo  {journal} {Phys. Rev. D}\ }\textbf {\bibinfo {volume}
  {79}},\ \bibinfo {pages} {115003} (\bibinfo {year} {2009}{\natexlab{b}})},\
  \Eprint {http://arxiv.org/abs/0903.4818} {arXiv:0903.4818 [hep-ph]}
  \BibitemShut {NoStop}%
\bibitem [{\citenamefont {Ballon-Bayona}\ \emph {et~al.}(2017)\citenamefont
  {Ballon-Bayona}, \citenamefont {Krein},\ and\ \citenamefont
  {Miller}}]{Ballon-Bayona:2017bwk}%
  \BibitemOpen
  \bibfield  {author} {\bibinfo {author} {\bibfnamefont {A.}~\bibnamefont
  {Ballon-Bayona}}, \bibinfo {author} {\bibfnamefont {G.}~\bibnamefont
  {Krein}}, \ and\ \bibinfo {author} {\bibfnamefont {C.}~\bibnamefont
  {Miller}},\ }\href {\doibase 10.1103/PhysRevD.96.014017} {\bibfield
  {journal} {\bibinfo  {journal} {Phys. Rev. D}\ }\textbf {\bibinfo {volume}
  {96}},\ \bibinfo {pages} {014017} (\bibinfo {year} {2017})},\ \Eprint
  {http://arxiv.org/abs/1702.08417} {arXiv:1702.08417 [hep-ph]} \BibitemShut
  {NoStop}%
\bibitem [{\citenamefont {Kumano}\ \emph {et~al.}(2018)\citenamefont {Kumano},
  \citenamefont {Song},\ and\ \citenamefont {Teryaev}}]{Kumano:2017lhr}%
  \BibitemOpen
  \bibfield  {author} {\bibinfo {author} {\bibfnamefont {S.}~\bibnamefont
  {Kumano}}, \bibinfo {author} {\bibfnamefont {Q.-T.}\ \bibnamefont {Song}}, \
  and\ \bibinfo {author} {\bibfnamefont {O.~V.}\ \bibnamefont {Teryaev}},\
  }\href {\doibase 10.1103/PhysRevD.97.014020} {\bibfield  {journal} {\bibinfo
  {journal} {Phys. Rev. D}\ }\textbf {\bibinfo {volume} {97}},\ \bibinfo
  {pages} {014020} (\bibinfo {year} {2018})},\ \Eprint
  {http://arxiv.org/abs/1711.08088} {arXiv:1711.08088 [hep-ph]} \BibitemShut
  {NoStop}%
\bibitem [{\citenamefont {Polyakov}\ and\ \citenamefont
  {Son}(2018)}]{Polyakov:2018exb}%
  \BibitemOpen
  \bibfield  {author} {\bibinfo {author} {\bibfnamefont {M.~V.}\ \bibnamefont
  {Polyakov}}\ and\ \bibinfo {author} {\bibfnamefont {H.-D.}\ \bibnamefont
  {Son}},\ }\href {\doibase 10.1007/JHEP09(2018)156} {\bibfield  {journal}
  {\bibinfo  {journal} {JHEP}\ }\textbf {\bibinfo {volume} {09}},\ \bibinfo
  {pages} {156} (\bibinfo {year} {2018})},\ \Eprint
  {http://arxiv.org/abs/1808.00155} {arXiv:1808.00155 [hep-ph]} \BibitemShut
  {NoStop}%
\bibitem [{\citenamefont {Freese}\ \emph {et~al.}(2019)\citenamefont {Freese},
  \citenamefont {Freese}, \citenamefont {Clo\"et},\ and\ \citenamefont
  {Clo\"et}}]{Freese:2019bhb}%
  \BibitemOpen
  \bibfield  {author} {\bibinfo {author} {\bibfnamefont {A.}~\bibnamefont
  {Freese}}, \bibinfo {author} {\bibfnamefont {A.}~\bibnamefont {Freese}},
  \bibinfo {author} {\bibfnamefont {I.~C.}\ \bibnamefont {Clo\"et}}, \ and\
  \bibinfo {author} {\bibfnamefont {I.~C.}\ \bibnamefont {Clo\"et}},\ }\href
  {\doibase 10.1103/PhysRevC.100.015201} {\bibfield  {journal} {\bibinfo
  {journal} {Phys. Rev. C}\ }\textbf {\bibinfo {volume} {100}},\ \bibinfo
  {pages} {015201} (\bibinfo {year} {2019})},\ \bibinfo {note} {[Erratum:
  Phys.Rev.C 105, 059901 (2022)]},\ \Eprint {http://arxiv.org/abs/1903.09222}
  {arXiv:1903.09222 [nucl-th]} \BibitemShut {NoStop}%
\bibitem [{\citenamefont {Polyakov}\ and\ \citenamefont
  {Sun}(2019)}]{Polyakov:2019lbq}%
  \BibitemOpen
  \bibfield  {author} {\bibinfo {author} {\bibfnamefont {M.~V.}\ \bibnamefont
  {Polyakov}}\ and\ \bibinfo {author} {\bibfnamefont {B.-D.}\ \bibnamefont
  {Sun}},\ }\href {\doibase 10.1103/PhysRevD.100.036003} {\bibfield  {journal}
  {\bibinfo  {journal} {Phys. Rev. D}\ }\textbf {\bibinfo {volume} {100}},\
  \bibinfo {pages} {036003} (\bibinfo {year} {2019})},\ \Eprint
  {http://arxiv.org/abs/1903.02738} {arXiv:1903.02738 [hep-ph]} \BibitemShut
  {NoStop}%
\bibitem [{\citenamefont {Krutov}\ and\ \citenamefont
  {Troitsky}(2021)}]{Krutov:2020ewr}%
  \BibitemOpen
  \bibfield  {author} {\bibinfo {author} {\bibfnamefont {A.~F.}\ \bibnamefont
  {Krutov}}\ and\ \bibinfo {author} {\bibfnamefont {V.~E.}\ \bibnamefont
  {Troitsky}},\ }\href {\doibase 10.1103/PhysRevD.103.014029} {\bibfield
  {journal} {\bibinfo  {journal} {Phys. Rev. D}\ }\textbf {\bibinfo {volume}
  {103}},\ \bibinfo {pages} {014029} (\bibinfo {year} {2021})},\ \Eprint
  {http://arxiv.org/abs/2010.11640} {arXiv:2010.11640 [hep-ph]} \BibitemShut
  {NoStop}%
\bibitem [{\citenamefont {Shi}\ \emph {et~al.}(2020)\citenamefont {Shi},
  \citenamefont {Bednar}, \citenamefont {Clo\"et},\ and\ \citenamefont
  {Freese}}]{Shi:2020pqe}%
  \BibitemOpen
  \bibfield  {author} {\bibinfo {author} {\bibfnamefont {C.}~\bibnamefont
  {Shi}}, \bibinfo {author} {\bibfnamefont {K.}~\bibnamefont {Bednar}},
  \bibinfo {author} {\bibfnamefont {I.~C.}\ \bibnamefont {Clo\"et}}, \ and\
  \bibinfo {author} {\bibfnamefont {A.}~\bibnamefont {Freese}},\ }\href
  {\doibase 10.1103/PhysRevD.101.074014} {\bibfield  {journal} {\bibinfo
  {journal} {Phys. Rev. D}\ }\textbf {\bibinfo {volume} {101}},\ \bibinfo
  {pages} {074014} (\bibinfo {year} {2020})},\ \Eprint
  {http://arxiv.org/abs/2003.03037} {arXiv:2003.03037 [hep-ph]} \BibitemShut
  {NoStop}%
\bibitem [{\citenamefont {Kim}\ and\ \citenamefont {Sun}(2021)}]{Kim:2020lrs}%
  \BibitemOpen
  \bibfield  {author} {\bibinfo {author} {\bibfnamefont {J.-Y.}\ \bibnamefont
  {Kim}}\ and\ \bibinfo {author} {\bibfnamefont {B.-D.}\ \bibnamefont {Sun}},\
  }\href {\doibase 10.1140/epjc/s10052-021-08852-z} {\bibfield  {journal}
  {\bibinfo  {journal} {Eur. Phys. J. C}\ }\textbf {\bibinfo {volume} {81}},\
  \bibinfo {pages} {85} (\bibinfo {year} {2021})},\ \Eprint
  {http://arxiv.org/abs/2011.00292} {arXiv:2011.00292 [hep-ph]} \BibitemShut
  {NoStop}%
\bibitem [{\citenamefont {Metz}\ \emph {et~al.}(2021)\citenamefont {Metz},
  \citenamefont {Pasquini},\ and\ \citenamefont {Rodini}}]{Metz:2021lqv}%
  \BibitemOpen
  \bibfield  {author} {\bibinfo {author} {\bibfnamefont {A.}~\bibnamefont
  {Metz}}, \bibinfo {author} {\bibfnamefont {B.}~\bibnamefont {Pasquini}}, \
  and\ \bibinfo {author} {\bibfnamefont {S.}~\bibnamefont {Rodini}},\ }\href
  {\doibase 10.1016/j.physletb.2021.136501} {\bibfield  {journal} {\bibinfo
  {journal} {Phys. Lett. B}\ }\textbf {\bibinfo {volume} {820}},\ \bibinfo
  {pages} {136501} (\bibinfo {year} {2021})},\ \Eprint
  {http://arxiv.org/abs/2104.04207} {arXiv:2104.04207 [hep-ph]} \BibitemShut
  {NoStop}%
\bibitem [{\citenamefont {Tong}\ \emph {et~al.}(2021)\citenamefont {Tong},
  \citenamefont {Ma},\ and\ \citenamefont {Yuan}}]{Tong:2021ctu}%
  \BibitemOpen
  \bibfield  {author} {\bibinfo {author} {\bibfnamefont {X.-B.}\ \bibnamefont
  {Tong}}, \bibinfo {author} {\bibfnamefont {J.-P.}\ \bibnamefont {Ma}}, \ and\
  \bibinfo {author} {\bibfnamefont {F.}~\bibnamefont {Yuan}},\ }\href {\doibase
  10.1016/j.physletb.2021.136751} {\bibfield  {journal} {\bibinfo  {journal}
  {Phys. Lett. B}\ }\textbf {\bibinfo {volume} {823}},\ \bibinfo {pages}
  {136751} (\bibinfo {year} {2021})},\ \Eprint
  {http://arxiv.org/abs/2101.02395} {arXiv:2101.02395 [hep-ph]} \BibitemShut
  {NoStop}%
\bibitem [{\citenamefont {Mamo}\ and\ \citenamefont
  {Zahed}(2022)}]{Mamo:2022eui}%
  \BibitemOpen
  \bibfield  {author} {\bibinfo {author} {\bibfnamefont {K.~A.}\ \bibnamefont
  {Mamo}}\ and\ \bibinfo {author} {\bibfnamefont {I.}~\bibnamefont {Zahed}},\
  }\href {\doibase 10.1103/PhysRevD.106.086004} {\bibfield  {journal} {\bibinfo
   {journal} {Phys. Rev. D}\ }\textbf {\bibinfo {volume} {106}},\ \bibinfo
  {pages} {086004} (\bibinfo {year} {2022})},\ \Eprint
  {http://arxiv.org/abs/2204.08857} {arXiv:2204.08857 [hep-ph]} \BibitemShut
  {NoStop}%
\bibitem [{\citenamefont {Krutov}\ and\ \citenamefont
  {Troitsky}(2022)}]{Krutov:2022zgg}%
  \BibitemOpen
  \bibfield  {author} {\bibinfo {author} {\bibfnamefont {A.~F.}\ \bibnamefont
  {Krutov}}\ and\ \bibinfo {author} {\bibfnamefont {V.~E.}\ \bibnamefont
  {Troitsky}},\ }\href {\doibase 10.1103/PhysRevD.106.054013} {\bibfield
  {journal} {\bibinfo  {journal} {Phys. Rev. D}\ }\textbf {\bibinfo {volume}
  {106}},\ \bibinfo {pages} {054013} (\bibinfo {year} {2022})},\ \Eprint
  {http://arxiv.org/abs/2201.04991} {arXiv:2201.04991 [hep-ph]} \BibitemShut
  {NoStop}%
\bibitem [{\citenamefont {Hackett}\ \emph {et~al.}(2024)\citenamefont
  {Hackett}, \citenamefont {Pefkou},\ and\ \citenamefont
  {Shanahan}}]{Hackett:2023rif}%
  \BibitemOpen
  \bibfield  {author} {\bibinfo {author} {\bibfnamefont {D.~C.}\ \bibnamefont
  {Hackett}}, \bibinfo {author} {\bibfnamefont {D.~A.}\ \bibnamefont {Pefkou}},
  \ and\ \bibinfo {author} {\bibfnamefont {P.~E.}\ \bibnamefont {Shanahan}},\
  }\href {\doibase 10.1103/PhysRevLett.132.251904} {\bibfield  {journal}
  {\bibinfo  {journal} {Phys. Rev. Lett.}\ }\textbf {\bibinfo {volume} {132}},\
  \bibinfo {pages} {251904} (\bibinfo {year} {2024})},\ \Eprint
  {http://arxiv.org/abs/2310.08484} {arXiv:2310.08484 [hep-lat]} \BibitemShut
  {NoStop}%
\bibitem [{\citenamefont {Guo}\ \emph {et~al.}(2023{\natexlab{a}})\citenamefont
  {Guo}, \citenamefont {Ji},\ and\ \citenamefont {Yuan}}]{Guo:2023qgu}%
  \BibitemOpen
  \bibfield  {author} {\bibinfo {author} {\bibfnamefont {Y.}~\bibnamefont
  {Guo}}, \bibinfo {author} {\bibfnamefont {X.}~\bibnamefont {Ji}}, \ and\
  \bibinfo {author} {\bibfnamefont {F.}~\bibnamefont {Yuan}},\ }\href@noop {}
  {\  (\bibinfo {year} {2023}{\natexlab{a}})},\ \Eprint
  {http://arxiv.org/abs/2308.13006} {arXiv:2308.13006 [hep-ph]} \BibitemShut
  {NoStop}%
\bibitem [{\citenamefont {Hackett}\ \emph {et~al.}(2023)\citenamefont
  {Hackett}, \citenamefont {Oare}, \citenamefont {Pefkou},\ and\ \citenamefont
  {Shanahan}}]{Hackett:2023nkr}%
  \BibitemOpen
  \bibfield  {author} {\bibinfo {author} {\bibfnamefont {D.~C.}\ \bibnamefont
  {Hackett}}, \bibinfo {author} {\bibfnamefont {P.~R.}\ \bibnamefont {Oare}},
  \bibinfo {author} {\bibfnamefont {D.~A.}\ \bibnamefont {Pefkou}}, \ and\
  \bibinfo {author} {\bibfnamefont {P.~E.}\ \bibnamefont {Shanahan}},\ }\href
  {\doibase 10.1103/PhysRevD.108.114504} {\bibfield  {journal} {\bibinfo
  {journal} {Phys. Rev. D}\ }\textbf {\bibinfo {volume} {108}},\ \bibinfo
  {pages} {114504} (\bibinfo {year} {2023})},\ \Eprint
  {http://arxiv.org/abs/2307.11707} {arXiv:2307.11707 [hep-lat]} \BibitemShut
  {NoStop}%
\bibitem [{\citenamefont {Guo}\ \emph {et~al.}(2023{\natexlab{b}})\citenamefont
  {Guo}, \citenamefont {Ji}, \citenamefont {Liu},\ and\ \citenamefont
  {Yang}}]{Guo:2023pqw}%
  \BibitemOpen
  \bibfield  {author} {\bibinfo {author} {\bibfnamefont {Y.}~\bibnamefont
  {Guo}}, \bibinfo {author} {\bibfnamefont {X.}~\bibnamefont {Ji}}, \bibinfo
  {author} {\bibfnamefont {Y.}~\bibnamefont {Liu}}, \ and\ \bibinfo {author}
  {\bibfnamefont {J.}~\bibnamefont {Yang}},\ }\href {\doibase
  10.1103/PhysRevD.108.034003} {\bibfield  {journal} {\bibinfo  {journal}
  {Phys. Rev. D}\ }\textbf {\bibinfo {volume} {108}},\ \bibinfo {pages}
  {034003} (\bibinfo {year} {2023}{\natexlab{b}})},\ \Eprint
  {http://arxiv.org/abs/2305.06992} {arXiv:2305.06992 [hep-ph]} \BibitemShut
  {NoStop}%
\bibitem [{\citenamefont {Hatta}(2024)}]{Hatta:2023fqc}%
  \BibitemOpen
  \bibfield  {author} {\bibinfo {author} {\bibfnamefont {Y.}~\bibnamefont
  {Hatta}},\ }\href {\doibase 10.1103/PhysRevD.109.L051502} {\bibfield
  {journal} {\bibinfo  {journal} {Phys. Rev. D}\ }\textbf {\bibinfo {volume}
  {109}},\ \bibinfo {pages} {L051502} (\bibinfo {year} {2024})},\ \Eprint
  {http://arxiv.org/abs/2311.14470} {arXiv:2311.14470 [hep-ph]} \BibitemShut
  {NoStop}%
\bibitem [{\citenamefont {Xu}\ \emph {et~al.}(2024)\citenamefont {Xu},
  \citenamefont {Ding}, \citenamefont {Raya}, \citenamefont {Roberts},
  \citenamefont {Rodr\'\i{}guez-Quintero},\ and\ \citenamefont
  {Schmidt}}]{Xu:2023izo}%
  \BibitemOpen
  \bibfield  {author} {\bibinfo {author} {\bibfnamefont {Y.-Z.}\ \bibnamefont
  {Xu}}, \bibinfo {author} {\bibfnamefont {M.}~\bibnamefont {Ding}}, \bibinfo
  {author} {\bibfnamefont {K.}~\bibnamefont {Raya}}, \bibinfo {author}
  {\bibfnamefont {C.~D.}\ \bibnamefont {Roberts}}, \bibinfo {author}
  {\bibfnamefont {J.}~\bibnamefont {Rodr\'\i{}guez-Quintero}}, \ and\ \bibinfo
  {author} {\bibfnamefont {S.~M.}\ \bibnamefont {Schmidt}},\ }\href {\doibase
  10.1140/epjc/s10052-024-12518-x} {\bibfield  {journal} {\bibinfo  {journal}
  {Eur. Phys. J. C}\ }\textbf {\bibinfo {volume} {84}},\ \bibinfo {pages} {191}
  (\bibinfo {year} {2024})},\ \Eprint {http://arxiv.org/abs/2311.14832}
  {arXiv:2311.14832 [hep-ph]} \BibitemShut {NoStop}%
\bibitem [{\citenamefont {Fujii}\ \emph {et~al.}(2024)\citenamefont {Fujii},
  \citenamefont {Iwanaka},\ and\ \citenamefont {Tanaka}}]{Fujii:2024rqd}%
  \BibitemOpen
  \bibfield  {author} {\bibinfo {author} {\bibfnamefont {D.}~\bibnamefont
  {Fujii}}, \bibinfo {author} {\bibfnamefont {A.}~\bibnamefont {Iwanaka}}, \
  and\ \bibinfo {author} {\bibfnamefont {M.}~\bibnamefont {Tanaka}},\ }\href
  {\doibase 10.1103/PhysRevD.110.L091501} {\bibfield  {journal} {\bibinfo
  {journal} {Phys. Rev. D}\ }\textbf {\bibinfo {volume} {110}},\ \bibinfo
  {pages} {L091501} (\bibinfo {year} {2024})},\ \Eprint
  {http://arxiv.org/abs/2407.21113} {arXiv:2407.21113 [hep-ph]} \BibitemShut
  {NoStop}%
\bibitem [{\citenamefont {Aharony}\ \emph {et~al.}(2000)\citenamefont
  {Aharony}, \citenamefont {Gubser}, \citenamefont {Maldacena}, \citenamefont
  {Ooguri},\ and\ \citenamefont {Oz}}]{Aharony:1999ti}%
  \BibitemOpen
  \bibfield  {author} {\bibinfo {author} {\bibfnamefont {O.}~\bibnamefont
  {Aharony}}, \bibinfo {author} {\bibfnamefont {S.~S.}\ \bibnamefont {Gubser}},
  \bibinfo {author} {\bibfnamefont {J.~M.}\ \bibnamefont {Maldacena}}, \bibinfo
  {author} {\bibfnamefont {H.}~\bibnamefont {Ooguri}}, \ and\ \bibinfo {author}
  {\bibfnamefont {Y.}~\bibnamefont {Oz}},\ }\href {\doibase
  10.1016/S0370-1573(99)00083-6} {\bibfield  {journal} {\bibinfo  {journal}
  {Phys. Rept.}\ }\textbf {\bibinfo {volume} {323}},\ \bibinfo {pages} {183}
  (\bibinfo {year} {2000})},\ \Eprint {http://arxiv.org/abs/hep-th/9905111}
  {arXiv:hep-th/9905111} \BibitemShut {NoStop}%
\bibitem [{\citenamefont {Kruczenski}\ \emph {et~al.}(2004)\citenamefont
  {Kruczenski}, \citenamefont {Mateos}, \citenamefont {Myers},\ and\
  \citenamefont {Winters}}]{Kruczenski:2003uq}%
  \BibitemOpen
  \bibfield  {author} {\bibinfo {author} {\bibfnamefont {M.}~\bibnamefont
  {Kruczenski}}, \bibinfo {author} {\bibfnamefont {D.}~\bibnamefont {Mateos}},
  \bibinfo {author} {\bibfnamefont {R.~C.}\ \bibnamefont {Myers}}, \ and\
  \bibinfo {author} {\bibfnamefont {D.~J.}\ \bibnamefont {Winters}},\ }\href
  {\doibase 10.1088/1126-6708/2004/05/041} {\bibfield  {journal} {\bibinfo
  {journal} {JHEP}\ }\textbf {\bibinfo {volume} {05}},\ \bibinfo {pages} {041}
  (\bibinfo {year} {2004})},\ \Eprint {http://arxiv.org/abs/hep-th/0311270}
  {arXiv:hep-th/0311270} \BibitemShut {NoStop}%
\bibitem [{\citenamefont {Sakai}\ and\ \citenamefont
  {Sugimoto}(2005{\natexlab{a}})}]{Sakai:2004cn}%
  \BibitemOpen
  \bibfield  {author} {\bibinfo {author} {\bibfnamefont {T.}~\bibnamefont
  {Sakai}}\ and\ \bibinfo {author} {\bibfnamefont {S.}~\bibnamefont
  {Sugimoto}},\ }\href {\doibase 10.1143/PTP.113.843} {\bibfield  {journal}
  {\bibinfo  {journal} {Prog. Theor. Phys.}\ }\textbf {\bibinfo {volume}
  {113}},\ \bibinfo {pages} {843} (\bibinfo {year} {2005}{\natexlab{a}})},\
  \Eprint {http://arxiv.org/abs/hep-th/0412141} {arXiv:hep-th/0412141}
  \BibitemShut {NoStop}%
\bibitem [{\citenamefont {Erlich}\ \emph {et~al.}(2005)\citenamefont {Erlich},
  \citenamefont {Katz}, \citenamefont {Son},\ and\ \citenamefont
  {Stephanov}}]{Erlich:2005qh}%
  \BibitemOpen
  \bibfield  {author} {\bibinfo {author} {\bibfnamefont {J.}~\bibnamefont
  {Erlich}}, \bibinfo {author} {\bibfnamefont {E.}~\bibnamefont {Katz}},
  \bibinfo {author} {\bibfnamefont {D.~T.}\ \bibnamefont {Son}}, \ and\
  \bibinfo {author} {\bibfnamefont {M.~A.}\ \bibnamefont {Stephanov}},\ }\href
  {\doibase 10.1103/PhysRevLett.95.261602} {\bibfield  {journal} {\bibinfo
  {journal} {Phys. Rev. Lett.}\ }\textbf {\bibinfo {volume} {95}},\ \bibinfo
  {pages} {261602} (\bibinfo {year} {2005})},\ \Eprint
  {http://arxiv.org/abs/hep-ph/0501128} {arXiv:hep-ph/0501128} \BibitemShut
  {NoStop}%
\bibitem [{\citenamefont {de~Teramond}\ and\ \citenamefont
  {Brodsky}(2005)}]{deTeramond:2005su}%
  \BibitemOpen
  \bibfield  {author} {\bibinfo {author} {\bibfnamefont {G.~F.}\ \bibnamefont
  {de~Teramond}}\ and\ \bibinfo {author} {\bibfnamefont {S.~J.}\ \bibnamefont
  {Brodsky}},\ }\href {\doibase 10.1103/PhysRevLett.94.201601} {\bibfield
  {journal} {\bibinfo  {journal} {Phys. Rev. Lett.}\ }\textbf {\bibinfo
  {volume} {94}},\ \bibinfo {pages} {201601} (\bibinfo {year} {2005})},\
  \Eprint {http://arxiv.org/abs/hep-th/0501022} {arXiv:hep-th/0501022}
  \BibitemShut {NoStop}%
\bibitem [{\citenamefont {Sakai}\ and\ \citenamefont
  {Sugimoto}(2005{\natexlab{b}})}]{Sakai:2005yt}%
  \BibitemOpen
  \bibfield  {author} {\bibinfo {author} {\bibfnamefont {T.}~\bibnamefont
  {Sakai}}\ and\ \bibinfo {author} {\bibfnamefont {S.}~\bibnamefont
  {Sugimoto}},\ }\href {\doibase 10.1143/PTP.114.1083} {\bibfield  {journal}
  {\bibinfo  {journal} {Prog. Theor. Phys.}\ }\textbf {\bibinfo {volume}
  {114}},\ \bibinfo {pages} {1083} (\bibinfo {year} {2005}{\natexlab{b}})},\
  \Eprint {http://arxiv.org/abs/hep-th/0507073} {arXiv:hep-th/0507073}
  \BibitemShut {NoStop}%
\bibitem [{\citenamefont {Brodsky}\ and\ \citenamefont
  {de~Teramond}(2006)}]{Brodsky:2006uqa}%
  \BibitemOpen
  \bibfield  {author} {\bibinfo {author} {\bibfnamefont {S.~J.}\ \bibnamefont
  {Brodsky}}\ and\ \bibinfo {author} {\bibfnamefont {G.~F.}\ \bibnamefont
  {de~Teramond}},\ }\href {\doibase 10.1103/PhysRevLett.96.201601} {\bibfield
  {journal} {\bibinfo  {journal} {Phys. Rev. Lett.}\ }\textbf {\bibinfo
  {volume} {96}},\ \bibinfo {pages} {201601} (\bibinfo {year} {2006})},\
  \Eprint {http://arxiv.org/abs/hep-ph/0602252} {arXiv:hep-ph/0602252}
  \BibitemShut {NoStop}%
\bibitem [{\citenamefont {Veneziano}(1968)}]{Veneziano:1968yb}%
  \BibitemOpen
  \bibfield  {author} {\bibinfo {author} {\bibfnamefont {G.}~\bibnamefont
  {Veneziano}},\ }\href {\doibase 10.1007/BF02824451} {\bibfield  {journal}
  {\bibinfo  {journal} {Nuovo Cim. A}\ }\textbf {\bibinfo {volume} {57}},\
  \bibinfo {pages} {190} (\bibinfo {year} {1968})}\BibitemShut {NoStop}%
\bibitem [{\citenamefont {Maldacena}(1998)}]{Maldacena:1997re}%
  \BibitemOpen
  \bibfield  {author} {\bibinfo {author} {\bibfnamefont {J.~M.}\ \bibnamefont
  {Maldacena}},\ }\href {\doibase 10.4310/ATMP.1998.v2.n2.a1} {\bibfield
  {journal} {\bibinfo  {journal} {Adv. Theor. Math. Phys.}\ }\textbf {\bibinfo
  {volume} {2}},\ \bibinfo {pages} {231} (\bibinfo {year} {1998})},\ \Eprint
  {http://arxiv.org/abs/hep-th/9711200} {arXiv:hep-th/9711200} \BibitemShut
  {NoStop}%
\bibitem [{\citenamefont {Gubser}\ \emph {et~al.}(1998)\citenamefont {Gubser},
  \citenamefont {Klebanov},\ and\ \citenamefont {Polyakov}}]{Gubser:1998bc}%
  \BibitemOpen
  \bibfield  {author} {\bibinfo {author} {\bibfnamefont {S.~S.}\ \bibnamefont
  {Gubser}}, \bibinfo {author} {\bibfnamefont {I.~R.}\ \bibnamefont
  {Klebanov}}, \ and\ \bibinfo {author} {\bibfnamefont {A.~M.}\ \bibnamefont
  {Polyakov}},\ }\href {\doibase 10.1016/S0370-2693(98)00377-3} {\bibfield
  {journal} {\bibinfo  {journal} {Phys. Lett. B}\ }\textbf {\bibinfo {volume}
  {428}},\ \bibinfo {pages} {105} (\bibinfo {year} {1998})},\ \Eprint
  {http://arxiv.org/abs/hep-th/9802109} {arXiv:hep-th/9802109} \BibitemShut
  {NoStop}%
\bibitem [{\citenamefont {Witten}(1998)}]{Witten:1998qj}%
  \BibitemOpen
  \bibfield  {author} {\bibinfo {author} {\bibfnamefont {E.}~\bibnamefont
  {Witten}},\ }\href {\doibase 10.4310/ATMP.1998.v2.n2.a2} {\bibfield
  {journal} {\bibinfo  {journal} {Adv. Theor. Math. Phys.}\ }\textbf {\bibinfo
  {volume} {2}},\ \bibinfo {pages} {253} (\bibinfo {year} {1998})},\ \Eprint
  {http://arxiv.org/abs/hep-th/9802150} {arXiv:hep-th/9802150} \BibitemShut
  {NoStop}%
\bibitem [{\citenamefont {Callan}\ \emph {et~al.}(1978)\citenamefont {Callan},
  \citenamefont {Dashen},\ and\ \citenamefont {Gross}}]{Callan:1977gz}%
  \BibitemOpen
  \bibfield  {author} {\bibinfo {author} {\bibfnamefont {C.~G.}\ \bibnamefont
  {Callan}, \bibfnamefont {Jr.}}, \bibinfo {author} {\bibfnamefont {R.~F.}\
  \bibnamefont {Dashen}}, \ and\ \bibinfo {author} {\bibfnamefont {D.~J.}\
  \bibnamefont {Gross}},\ }\href {\doibase 10.1103/PhysRevD.17.2717} {\bibfield
   {journal} {\bibinfo  {journal} {Phys. Rev. D}\ }\textbf {\bibinfo {volume}
  {17}},\ \bibinfo {pages} {2717} (\bibinfo {year} {1978})}\BibitemShut
  {NoStop}%
\bibitem [{\citenamefont {Brodsky}\ and\ \citenamefont
  {de~Teramond}(2008)}]{Brodsky:2007hb}%
  \BibitemOpen
  \bibfield  {author} {\bibinfo {author} {\bibfnamefont {S.~J.}\ \bibnamefont
  {Brodsky}}\ and\ \bibinfo {author} {\bibfnamefont {G.~F.}\ \bibnamefont
  {de~Teramond}},\ }\href {\doibase 10.1103/PhysRevD.77.056007} {\bibfield
  {journal} {\bibinfo  {journal} {Phys. Rev. D}\ }\textbf {\bibinfo {volume}
  {77}},\ \bibinfo {pages} {056007} (\bibinfo {year} {2008})},\ \Eprint
  {http://arxiv.org/abs/0707.3859} {arXiv:0707.3859 [hep-ph]} \BibitemShut
  {NoStop}%
\bibitem [{\citenamefont {Kwee}\ and\ \citenamefont
  {Lebed}(2008{\natexlab{a}})}]{Kwee:2007nq}%
  \BibitemOpen
  \bibfield  {author} {\bibinfo {author} {\bibfnamefont {H.~J.}\ \bibnamefont
  {Kwee}}\ and\ \bibinfo {author} {\bibfnamefont {R.~F.}\ \bibnamefont
  {Lebed}},\ }\href {\doibase 10.1103/PhysRevD.77.115007} {\bibfield  {journal}
  {\bibinfo  {journal} {Phys. Rev. D}\ }\textbf {\bibinfo {volume} {77}},\
  \bibinfo {pages} {115007} (\bibinfo {year} {2008}{\natexlab{a}})},\ \Eprint
  {http://arxiv.org/abs/0712.1811} {arXiv:0712.1811 [hep-ph]} \BibitemShut
  {NoStop}%
\bibitem [{\citenamefont {Branz}\ \emph {et~al.}(2010)\citenamefont {Branz},
  \citenamefont {Gutsche}, \citenamefont {Lyubovitskij}, \citenamefont
  {Schmidt},\ and\ \citenamefont {Vega}}]{Branz:2010ub}%
  \BibitemOpen
  \bibfield  {author} {\bibinfo {author} {\bibfnamefont {T.}~\bibnamefont
  {Branz}}, \bibinfo {author} {\bibfnamefont {T.}~\bibnamefont {Gutsche}},
  \bibinfo {author} {\bibfnamefont {V.~E.}\ \bibnamefont {Lyubovitskij}},
  \bibinfo {author} {\bibfnamefont {I.}~\bibnamefont {Schmidt}}, \ and\
  \bibinfo {author} {\bibfnamefont {A.}~\bibnamefont {Vega}},\ }\href {\doibase
  10.1103/PhysRevD.82.074022} {\bibfield  {journal} {\bibinfo  {journal} {Phys.
  Rev. D}\ }\textbf {\bibinfo {volume} {82}},\ \bibinfo {pages} {074022}
  (\bibinfo {year} {2010})},\ \Eprint {http://arxiv.org/abs/1008.0268}
  {arXiv:1008.0268 [hep-ph]} \BibitemShut {NoStop}%
\bibitem [{\citenamefont {Gutsche}\ \emph {et~al.}(2012)\citenamefont
  {Gutsche}, \citenamefont {Lyubovitskij}, \citenamefont {Schmidt},\ and\
  \citenamefont {Vega}}]{Gutsche:2011vb}%
  \BibitemOpen
  \bibfield  {author} {\bibinfo {author} {\bibfnamefont {T.}~\bibnamefont
  {Gutsche}}, \bibinfo {author} {\bibfnamefont {V.~E.}\ \bibnamefont
  {Lyubovitskij}}, \bibinfo {author} {\bibfnamefont {I.}~\bibnamefont
  {Schmidt}}, \ and\ \bibinfo {author} {\bibfnamefont {A.}~\bibnamefont
  {Vega}},\ }\href {\doibase 10.1103/PhysRevD.85.076003} {\bibfield  {journal}
  {\bibinfo  {journal} {Phys. Rev. D}\ }\textbf {\bibinfo {volume} {85}},\
  \bibinfo {pages} {076003} (\bibinfo {year} {2012})},\ \Eprint
  {http://arxiv.org/abs/1108.0346} {arXiv:1108.0346 [hep-ph]} \BibitemShut
  {NoStop}%
\bibitem [{\citenamefont {Li}\ and\ \citenamefont {Huang}(2013)}]{Li:2013oda}%
  \BibitemOpen
  \bibfield  {author} {\bibinfo {author} {\bibfnamefont {D.}~\bibnamefont
  {Li}}\ and\ \bibinfo {author} {\bibfnamefont {M.}~\bibnamefont {Huang}},\
  }\href {\doibase 10.1007/JHEP11(2013)088} {\bibfield  {journal} {\bibinfo
  {journal} {JHEP}\ }\textbf {\bibinfo {volume} {11}},\ \bibinfo {pages} {088}
  (\bibinfo {year} {2013})},\ \Eprint {http://arxiv.org/abs/1303.6929}
  {arXiv:1303.6929 [hep-ph]} \BibitemShut {NoStop}%
\bibitem [{\citenamefont {Brodsky}\ \emph {et~al.}(2015)\citenamefont
  {Brodsky}, \citenamefont {de~Teramond}, \citenamefont {Dosch},\ and\
  \citenamefont {Erlich}}]{Brodsky:2014yha}%
  \BibitemOpen
  \bibfield  {author} {\bibinfo {author} {\bibfnamefont {S.~J.}\ \bibnamefont
  {Brodsky}}, \bibinfo {author} {\bibfnamefont {G.~F.}\ \bibnamefont
  {de~Teramond}}, \bibinfo {author} {\bibfnamefont {H.~G.}\ \bibnamefont
  {Dosch}}, \ and\ \bibinfo {author} {\bibfnamefont {J.}~\bibnamefont
  {Erlich}},\ }\href {\doibase 10.1016/j.physrep.2015.05.001} {\bibfield
  {journal} {\bibinfo  {journal} {Phys. Rept.}\ }\textbf {\bibinfo {volume}
  {584}},\ \bibinfo {pages} {1} (\bibinfo {year} {2015})},\ \Eprint
  {http://arxiv.org/abs/1407.8131} {arXiv:1407.8131 [hep-ph]} \BibitemShut
  {NoStop}%
\bibitem [{\citenamefont {Gutsche}\ \emph {et~al.}(2018)\citenamefont
  {Gutsche}, \citenamefont {Lyubovitskij},\ and\ \citenamefont
  {Schmidt}}]{Gutsche:2017lyu}%
  \BibitemOpen
  \bibfield  {author} {\bibinfo {author} {\bibfnamefont {T.}~\bibnamefont
  {Gutsche}}, \bibinfo {author} {\bibfnamefont {V.~E.}\ \bibnamefont
  {Lyubovitskij}}, \ and\ \bibinfo {author} {\bibfnamefont {I.}~\bibnamefont
  {Schmidt}},\ }\href {\doibase 10.1103/PhysRevD.97.054011} {\bibfield
  {journal} {\bibinfo  {journal} {Phys. Rev. D}\ }\textbf {\bibinfo {volume}
  {97}},\ \bibinfo {pages} {054011} (\bibinfo {year} {2018})},\ \Eprint
  {http://arxiv.org/abs/1712.08410} {arXiv:1712.08410 [hep-ph]} \BibitemShut
  {NoStop}%
\bibitem [{\citenamefont {Lyubovitskij}\ and\ \citenamefont
  {Schmidt}(2020)}]{Lyubovitskij:2020gjz}%
  \BibitemOpen
  \bibfield  {author} {\bibinfo {author} {\bibfnamefont {V.~E.}\ \bibnamefont
  {Lyubovitskij}}\ and\ \bibinfo {author} {\bibfnamefont {I.}~\bibnamefont
  {Schmidt}},\ }\href {\doibase 10.1103/PhysRevD.102.094008} {\bibfield
  {journal} {\bibinfo  {journal} {Phys. Rev. D}\ }\textbf {\bibinfo {volume}
  {102}},\ \bibinfo {pages} {094008} (\bibinfo {year} {2020})},\ \Eprint
  {http://arxiv.org/abs/2009.07115} {arXiv:2009.07115 [hep-ph]} \BibitemShut
  {NoStop}%
\bibitem [{\citenamefont {Chen}\ and\ \citenamefont
  {Huang}(2022)}]{Chen:2021wzj}%
  \BibitemOpen
  \bibfield  {author} {\bibinfo {author} {\bibfnamefont {Y.}~\bibnamefont
  {Chen}}\ and\ \bibinfo {author} {\bibfnamefont {M.}~\bibnamefont {Huang}},\
  }\href {\doibase 10.1103/PhysRevD.105.026021} {\bibfield  {journal} {\bibinfo
   {journal} {Phys. Rev. D}\ }\textbf {\bibinfo {volume} {105}},\ \bibinfo
  {pages} {026021} (\bibinfo {year} {2022})},\ \Eprint
  {http://arxiv.org/abs/2110.08215} {arXiv:2110.08215 [hep-ph]} \BibitemShut
  {NoStop}%
\bibitem [{\citenamefont {Zhang}\ \emph {et~al.}(2022)\citenamefont {Zhang},
  \citenamefont {Chen}, \citenamefont {Chen},\ and\ \citenamefont
  {Huang}}]{Zhang:2021itx}%
  \BibitemOpen
  \bibfield  {author} {\bibinfo {author} {\bibfnamefont {L.}~\bibnamefont
  {Zhang}}, \bibinfo {author} {\bibfnamefont {C.}~\bibnamefont {Chen}},
  \bibinfo {author} {\bibfnamefont {Y.}~\bibnamefont {Chen}}, \ and\ \bibinfo
  {author} {\bibfnamefont {M.}~\bibnamefont {Huang}},\ }\href {\doibase
  10.1103/PhysRevD.105.026020} {\bibfield  {journal} {\bibinfo  {journal}
  {Phys. Rev. D}\ }\textbf {\bibinfo {volume} {105}},\ \bibinfo {pages}
  {026020} (\bibinfo {year} {2022})},\ \Eprint
  {http://arxiv.org/abs/2106.10748} {arXiv:2106.10748 [hep-ph]} \BibitemShut
  {NoStop}%
\bibitem [{\citenamefont {Chen}\ \emph {et~al.}(2022)\citenamefont {Chen},
  \citenamefont {Li},\ and\ \citenamefont {Huang}}]{Chen:2022goa}%
  \BibitemOpen
  \bibfield  {author} {\bibinfo {author} {\bibfnamefont {Y.}~\bibnamefont
  {Chen}}, \bibinfo {author} {\bibfnamefont {D.}~\bibnamefont {Li}}, \ and\
  \bibinfo {author} {\bibfnamefont {M.}~\bibnamefont {Huang}},\ }\href
  {\doibase 10.1088/1572-9494/ac82ad} {\bibfield  {journal} {\bibinfo
  {journal} {Commun. Theor. Phys.}\ }\textbf {\bibinfo {volume} {74}},\
  \bibinfo {pages} {097201} (\bibinfo {year} {2022})},\ \Eprint
  {http://arxiv.org/abs/2206.00917} {arXiv:2206.00917 [hep-ph]} \BibitemShut
  {NoStop}%
\bibitem [{\citenamefont {Gursoy}\ \emph {et~al.}(2011)\citenamefont {Gursoy},
  \citenamefont {Kiritsis}, \citenamefont {Mazzanti}, \citenamefont
  {Michalogiorgakis},\ and\ \citenamefont {Nitti}}]{Gursoy:2010fj}%
  \BibitemOpen
  \bibfield  {author} {\bibinfo {author} {\bibfnamefont {U.}~\bibnamefont
  {Gursoy}}, \bibinfo {author} {\bibfnamefont {E.}~\bibnamefont {Kiritsis}},
  \bibinfo {author} {\bibfnamefont {L.}~\bibnamefont {Mazzanti}}, \bibinfo
  {author} {\bibfnamefont {G.}~\bibnamefont {Michalogiorgakis}}, \ and\
  \bibinfo {author} {\bibfnamefont {F.}~\bibnamefont {Nitti}},\ }\href
  {\doibase 10.1007/978-3-642-04864-7_4} {\bibfield  {journal} {\bibinfo
  {journal} {Lect. Notes Phys.}\ }\textbf {\bibinfo {volume} {828}},\ \bibinfo
  {pages} {79} (\bibinfo {year} {2011})},\ \Eprint
  {http://arxiv.org/abs/1006.5461} {arXiv:1006.5461 [hep-th]} \BibitemShut
  {NoStop}%
\bibitem [{\citenamefont {Bergman}\ \emph {et~al.}(2007)\citenamefont
  {Bergman}, \citenamefont {Lifschytz},\ and\ \citenamefont
  {Lippert}}]{Bergman:2007wp}%
  \BibitemOpen
  \bibfield  {author} {\bibinfo {author} {\bibfnamefont {O.}~\bibnamefont
  {Bergman}}, \bibinfo {author} {\bibfnamefont {G.}~\bibnamefont {Lifschytz}},
  \ and\ \bibinfo {author} {\bibfnamefont {M.}~\bibnamefont {Lippert}},\ }\href
  {\doibase 10.1088/1126-6708/2007/11/056} {\bibfield  {journal} {\bibinfo
  {journal} {JHEP}\ }\textbf {\bibinfo {volume} {11}},\ \bibinfo {pages} {056}
  (\bibinfo {year} {2007})},\ \Eprint {http://arxiv.org/abs/0708.0326}
  {arXiv:0708.0326 [hep-th]} \BibitemShut {NoStop}%
\bibitem [{\citenamefont {Kwee}\ and\ \citenamefont
  {Lebed}(2008{\natexlab{b}})}]{Kwee:2007dd}%
  \BibitemOpen
  \bibfield  {author} {\bibinfo {author} {\bibfnamefont {H.~J.}\ \bibnamefont
  {Kwee}}\ and\ \bibinfo {author} {\bibfnamefont {R.~F.}\ \bibnamefont
  {Lebed}},\ }\href {\doibase 10.1088/1126-6708/2008/01/027} {\bibfield
  {journal} {\bibinfo  {journal} {JHEP}\ }\textbf {\bibinfo {volume} {01}},\
  \bibinfo {pages} {027} (\bibinfo {year} {2008}{\natexlab{b}})},\ \Eprint
  {http://arxiv.org/abs/0708.4054} {arXiv:0708.4054 [hep-ph]} \BibitemShut
  {NoStop}%
\bibitem [{\citenamefont {Grigoryan}\ and\ \citenamefont
  {Radyushkin}(2007{\natexlab{a}})}]{Grigoryan:2007wn}%
  \BibitemOpen
  \bibfield  {author} {\bibinfo {author} {\bibfnamefont {H.~R.}\ \bibnamefont
  {Grigoryan}}\ and\ \bibinfo {author} {\bibfnamefont {A.~V.}\ \bibnamefont
  {Radyushkin}},\ }\href {\doibase 10.1103/PhysRevD.76.115007} {\bibfield
  {journal} {\bibinfo  {journal} {Phys. Rev. D}\ }\textbf {\bibinfo {volume}
  {76}},\ \bibinfo {pages} {115007} (\bibinfo {year} {2007}{\natexlab{a}})},\
  \Eprint {http://arxiv.org/abs/0709.0500} {arXiv:0709.0500 [hep-ph]}
  \BibitemShut {NoStop}%
\bibitem [{\citenamefont {Grigoryan}\ and\ \citenamefont
  {Radyushkin}(2007{\natexlab{b}})}]{Grigoryan:2007vg}%
  \BibitemOpen
  \bibfield  {author} {\bibinfo {author} {\bibfnamefont {H.~R.}\ \bibnamefont
  {Grigoryan}}\ and\ \bibinfo {author} {\bibfnamefont {A.~V.}\ \bibnamefont
  {Radyushkin}},\ }\href {\doibase 10.1016/j.physletb.2007.05.044} {\bibfield
  {journal} {\bibinfo  {journal} {Phys. Lett. B}\ }\textbf {\bibinfo {volume}
  {650}},\ \bibinfo {pages} {421} (\bibinfo {year} {2007}{\natexlab{b}})},\
  \Eprint {http://arxiv.org/abs/hep-ph/0703069} {arXiv:hep-ph/0703069}
  \BibitemShut {NoStop}%
\bibitem [{\citenamefont {Abidin}\ and\ \citenamefont
  {Hutauruk}(2019)}]{Abidin:2019xwu}%
  \BibitemOpen
  \bibfield  {author} {\bibinfo {author} {\bibfnamefont {Z.}~\bibnamefont
  {Abidin}}\ and\ \bibinfo {author} {\bibfnamefont {P.~T.~P.}\ \bibnamefont
  {Hutauruk}},\ }\href {\doibase 10.1103/PhysRevD.100.054026} {\bibfield
  {journal} {\bibinfo  {journal} {Phys. Rev. D}\ }\textbf {\bibinfo {volume}
  {100}},\ \bibinfo {pages} {054026} (\bibinfo {year} {2019})},\ \Eprint
  {http://arxiv.org/abs/1905.08953} {arXiv:1905.08953 [hep-ph]} \BibitemShut
  {NoStop}%
\bibitem [{\citenamefont {Ahmed}\ \emph {et~al.}(2023)\citenamefont {Ahmed},
  \citenamefont {Chen},\ and\ \citenamefont {Huang}}]{Ahmed:2023zkk}%
  \BibitemOpen
  \bibfield  {author} {\bibinfo {author} {\bibfnamefont {H.~A.}\ \bibnamefont
  {Ahmed}}, \bibinfo {author} {\bibfnamefont {Y.}~\bibnamefont {Chen}}, \ and\
  \bibinfo {author} {\bibfnamefont {M.}~\bibnamefont {Huang}},\ }\href
  {\doibase 10.1103/PhysRevD.108.086034} {\bibfield  {journal} {\bibinfo
  {journal} {Phys. Rev. D}\ }\textbf {\bibinfo {volume} {108}},\ \bibinfo
  {pages} {086034} (\bibinfo {year} {2023})},\ \Eprint
  {http://arxiv.org/abs/2308.14975} {arXiv:2308.14975 [hep-ph]} \BibitemShut
  {NoStop}%
\bibitem [{\citenamefont {Polchinski}\ and\ \citenamefont
  {Strassler}(2002)}]{Polchinski:2001tt}%
  \BibitemOpen
  \bibfield  {author} {\bibinfo {author} {\bibfnamefont {J.}~\bibnamefont
  {Polchinski}}\ and\ \bibinfo {author} {\bibfnamefont {M.~J.}\ \bibnamefont
  {Strassler}},\ }\href {\doibase 10.1103/PhysRevLett.88.031601} {\bibfield
  {journal} {\bibinfo  {journal} {Phys. Rev. Lett.}\ }\textbf {\bibinfo
  {volume} {88}},\ \bibinfo {pages} {031601} (\bibinfo {year} {2002})},\
  \Eprint {http://arxiv.org/abs/hep-th/0109174} {arXiv:hep-th/0109174}
  \BibitemShut {NoStop}%
\bibitem [{\citenamefont {Polchinski}\ and\ \citenamefont
  {Strassler}(2003)}]{Polchinski:2002jw}%
  \BibitemOpen
  \bibfield  {author} {\bibinfo {author} {\bibfnamefont {J.}~\bibnamefont
  {Polchinski}}\ and\ \bibinfo {author} {\bibfnamefont {M.~J.}\ \bibnamefont
  {Strassler}},\ }\href {\doibase 10.1088/1126-6708/2003/05/012} {\bibfield
  {journal} {\bibinfo  {journal} {JHEP}\ }\textbf {\bibinfo {volume} {05}},\
  \bibinfo {pages} {012} (\bibinfo {year} {2003})},\ \Eprint
  {http://arxiv.org/abs/hep-th/0209211} {arXiv:hep-th/0209211} \BibitemShut
  {NoStop}%
\bibitem [{\citenamefont {Brower}\ \emph {et~al.}(2007)\citenamefont {Brower},
  \citenamefont {Polchinski}, \citenamefont {Strassler},\ and\ \citenamefont
  {Tan}}]{Brower:2006ea}%
  \BibitemOpen
  \bibfield  {author} {\bibinfo {author} {\bibfnamefont {R.~C.}\ \bibnamefont
  {Brower}}, \bibinfo {author} {\bibfnamefont {J.}~\bibnamefont {Polchinski}},
  \bibinfo {author} {\bibfnamefont {M.~J.}\ \bibnamefont {Strassler}}, \ and\
  \bibinfo {author} {\bibfnamefont {C.-I.}\ \bibnamefont {Tan}},\ }\href
  {\doibase 10.1088/1126-6708/2007/12/005} {\bibfield  {journal} {\bibinfo
  {journal} {JHEP}\ }\textbf {\bibinfo {volume} {12}},\ \bibinfo {pages} {005}
  (\bibinfo {year} {2007})},\ \Eprint {http://arxiv.org/abs/hep-th/0603115}
  {arXiv:hep-th/0603115} \BibitemShut {NoStop}%
\bibitem [{\citenamefont {Hatta}\ \emph {et~al.}(2008)\citenamefont {Hatta},
  \citenamefont {Iancu},\ and\ \citenamefont {Mueller}}]{Hatta:2007he}%
  \BibitemOpen
  \bibfield  {author} {\bibinfo {author} {\bibfnamefont {Y.}~\bibnamefont
  {Hatta}}, \bibinfo {author} {\bibfnamefont {E.}~\bibnamefont {Iancu}}, \ and\
  \bibinfo {author} {\bibfnamefont {A.~H.}\ \bibnamefont {Mueller}},\ }\href
  {\doibase 10.1088/1126-6708/2008/01/026} {\bibfield  {journal} {\bibinfo
  {journal} {JHEP}\ }\textbf {\bibinfo {volume} {01}},\ \bibinfo {pages} {026}
  (\bibinfo {year} {2008})},\ \Eprint {http://arxiv.org/abs/0710.2148}
  {arXiv:0710.2148 [hep-th]} \BibitemShut {NoStop}%
\bibitem [{\citenamefont {Pire}\ \emph {et~al.}(2008)\citenamefont {Pire},
  \citenamefont {Roiesnel}, \citenamefont {Szymanowski},\ and\ \citenamefont
  {Wallon}}]{Pire:2008zf}%
  \BibitemOpen
  \bibfield  {author} {\bibinfo {author} {\bibfnamefont {B.}~\bibnamefont
  {Pire}}, \bibinfo {author} {\bibfnamefont {C.}~\bibnamefont {Roiesnel}},
  \bibinfo {author} {\bibfnamefont {L.}~\bibnamefont {Szymanowski}}, \ and\
  \bibinfo {author} {\bibfnamefont {S.}~\bibnamefont {Wallon}},\ }\href
  {\doibase 10.1016/j.physletb.2008.10.026} {\bibfield  {journal} {\bibinfo
  {journal} {Phys. Lett. B}\ }\textbf {\bibinfo {volume} {670}},\ \bibinfo
  {pages} {84} (\bibinfo {year} {2008})},\ \Eprint
  {http://arxiv.org/abs/0805.4346} {arXiv:0805.4346 [hep-ph]} \BibitemShut
  {NoStop}%
\bibitem [{\citenamefont {Domokos}\ \emph {et~al.}(2009)\citenamefont
  {Domokos}, \citenamefont {Harvey},\ and\ \citenamefont
  {Mann}}]{Domokos:2009hm}%
  \BibitemOpen
  \bibfield  {author} {\bibinfo {author} {\bibfnamefont {S.~K.}\ \bibnamefont
  {Domokos}}, \bibinfo {author} {\bibfnamefont {J.~A.}\ \bibnamefont {Harvey}},
  \ and\ \bibinfo {author} {\bibfnamefont {N.}~\bibnamefont {Mann}},\ }\href
  {\doibase 10.1103/PhysRevD.80.126015} {\bibfield  {journal} {\bibinfo
  {journal} {Phys. Rev. D}\ }\textbf {\bibinfo {volume} {80}},\ \bibinfo
  {pages} {126015} (\bibinfo {year} {2009})},\ \Eprint
  {http://arxiv.org/abs/0907.1084} {arXiv:0907.1084 [hep-ph]} \BibitemShut
  {NoStop}%
\bibitem [{\citenamefont {Domokos}\ \emph {et~al.}(2010)\citenamefont
  {Domokos}, \citenamefont {Harvey},\ and\ \citenamefont
  {Mann}}]{Domokos:2010ma}%
  \BibitemOpen
  \bibfield  {author} {\bibinfo {author} {\bibfnamefont {S.~K.}\ \bibnamefont
  {Domokos}}, \bibinfo {author} {\bibfnamefont {J.~A.}\ \bibnamefont {Harvey}},
  \ and\ \bibinfo {author} {\bibfnamefont {N.}~\bibnamefont {Mann}},\ }\href
  {\doibase 10.1103/PhysRevD.82.106007} {\bibfield  {journal} {\bibinfo
  {journal} {Phys. Rev. D}\ }\textbf {\bibinfo {volume} {82}},\ \bibinfo
  {pages} {106007} (\bibinfo {year} {2010})},\ \Eprint
  {http://arxiv.org/abs/1008.2963} {arXiv:1008.2963 [hep-th]} \BibitemShut
  {NoStop}%
\bibitem [{\citenamefont {Marquet}\ \emph {et~al.}(2010)\citenamefont
  {Marquet}, \citenamefont {Roiesnel},\ and\ \citenamefont
  {Wallon}}]{Marquet:2010sf}%
  \BibitemOpen
  \bibfield  {author} {\bibinfo {author} {\bibfnamefont {C.}~\bibnamefont
  {Marquet}}, \bibinfo {author} {\bibfnamefont {C.}~\bibnamefont {Roiesnel}}, \
  and\ \bibinfo {author} {\bibfnamefont {S.}~\bibnamefont {Wallon}},\ }\href
  {\doibase 10.1007/JHEP04(2010)051} {\bibfield  {journal} {\bibinfo  {journal}
  {JHEP}\ }\textbf {\bibinfo {volume} {04}},\ \bibinfo {pages} {051} (\bibinfo
  {year} {2010})},\ \Eprint {http://arxiv.org/abs/1002.0566} {arXiv:1002.0566
  [hep-ph]} \BibitemShut {NoStop}%
\bibitem [{\citenamefont {Watanabe}\ and\ \citenamefont
  {Suzuki}(2012)}]{Watanabe:2012uc}%
  \BibitemOpen
  \bibfield  {author} {\bibinfo {author} {\bibfnamefont {A.}~\bibnamefont
  {Watanabe}}\ and\ \bibinfo {author} {\bibfnamefont {K.}~\bibnamefont
  {Suzuki}},\ }\href {\doibase 10.1103/PhysRevD.86.035011} {\bibfield
  {journal} {\bibinfo  {journal} {Phys. Rev. D}\ }\textbf {\bibinfo {volume}
  {86}},\ \bibinfo {pages} {035011} (\bibinfo {year} {2012})},\ \Eprint
  {http://arxiv.org/abs/1206.0910} {arXiv:1206.0910 [hep-ph]} \BibitemShut
  {NoStop}%
\bibitem [{\citenamefont {Watanabe}\ and\ \citenamefont
  {Suzuki}(2014)}]{Watanabe:2013spa}%
  \BibitemOpen
  \bibfield  {author} {\bibinfo {author} {\bibfnamefont {A.}~\bibnamefont
  {Watanabe}}\ and\ \bibinfo {author} {\bibfnamefont {K.}~\bibnamefont
  {Suzuki}},\ }\href {\doibase 10.1103/PhysRevD.89.115015} {\bibfield
  {journal} {\bibinfo  {journal} {Phys. Rev. D}\ }\textbf {\bibinfo {volume}
  {89}},\ \bibinfo {pages} {115015} (\bibinfo {year} {2014})},\ \Eprint
  {http://arxiv.org/abs/1312.7114} {arXiv:1312.7114 [hep-ph]} \BibitemShut
  {NoStop}%
\bibitem [{\citenamefont {Watanabe}\ and\ \citenamefont
  {Li}(2015)}]{Watanabe:2015mia}%
  \BibitemOpen
  \bibfield  {author} {\bibinfo {author} {\bibfnamefont {A.}~\bibnamefont
  {Watanabe}}\ and\ \bibinfo {author} {\bibfnamefont {H.-n.}\ \bibnamefont
  {Li}},\ }\href {\doibase 10.1016/j.physletb.2015.10.069} {\bibfield
  {journal} {\bibinfo  {journal} {Phys. Lett. B}\ }\textbf {\bibinfo {volume}
  {751}},\ \bibinfo {pages} {321} (\bibinfo {year} {2015})},\ \Eprint
  {http://arxiv.org/abs/1502.03894} {arXiv:1502.03894 [hep-ph]} \BibitemShut
  {NoStop}%
\bibitem [{\citenamefont {Anderson}\ \emph {et~al.}(2017)\citenamefont
  {Anderson}, \citenamefont {Domokos},\ and\ \citenamefont
  {Mann}}]{Anderson:2016zon}%
  \BibitemOpen
  \bibfield  {author} {\bibinfo {author} {\bibfnamefont {N.}~\bibnamefont
  {Anderson}}, \bibinfo {author} {\bibfnamefont {S.}~\bibnamefont {Domokos}}, \
  and\ \bibinfo {author} {\bibfnamefont {N.}~\bibnamefont {Mann}},\ }\href
  {\doibase 10.1103/PhysRevD.96.046002} {\bibfield  {journal} {\bibinfo
  {journal} {Phys. Rev. D}\ }\textbf {\bibinfo {volume} {96}},\ \bibinfo
  {pages} {046002} (\bibinfo {year} {2017})},\ \Eprint
  {http://arxiv.org/abs/1612.07457} {arXiv:1612.07457 [hep-ph]} \BibitemShut
  {NoStop}%
\bibitem [{\citenamefont {Watanabe}\ and\ \citenamefont
  {Huang}(2019)}]{Watanabe:2018owy}%
  \BibitemOpen
  \bibfield  {author} {\bibinfo {author} {\bibfnamefont {A.}~\bibnamefont
  {Watanabe}}\ and\ \bibinfo {author} {\bibfnamefont {M.}~\bibnamefont
  {Huang}},\ }\href {\doibase 10.1016/j.physletb.2018.11.042} {\bibfield
  {journal} {\bibinfo  {journal} {Phys. Lett. B}\ }\textbf {\bibinfo {volume}
  {788}},\ \bibinfo {pages} {256} (\bibinfo {year} {2019})},\ \Eprint
  {http://arxiv.org/abs/1809.02515} {arXiv:1809.02515 [hep-ph]} \BibitemShut
  {NoStop}%
\bibitem [{\citenamefont {Burikham}\ and\ \citenamefont
  {Samart}(2019)}]{Burikham:2019zbo}%
  \BibitemOpen
  \bibfield  {author} {\bibinfo {author} {\bibfnamefont {P.}~\bibnamefont
  {Burikham}}\ and\ \bibinfo {author} {\bibfnamefont {D.}~\bibnamefont
  {Samart}},\ }\href {\doibase 10.1140/epjc/s10052-019-6957-3} {\bibfield
  {journal} {\bibinfo  {journal} {Eur. Phys. J. C}\ }\textbf {\bibinfo {volume}
  {79}},\ \bibinfo {pages} {452} (\bibinfo {year} {2019})},\ \Eprint
  {http://arxiv.org/abs/1902.05706} {arXiv:1902.05706 [hep-ph]} \BibitemShut
  {NoStop}%
\bibitem [{\citenamefont {Watanabe}\ \emph {et~al.}(2020)\citenamefont
  {Watanabe}, \citenamefont {Sawada},\ and\ \citenamefont
  {Huang}}]{Watanabe:2019zny}%
  \BibitemOpen
  \bibfield  {author} {\bibinfo {author} {\bibfnamefont {A.}~\bibnamefont
  {Watanabe}}, \bibinfo {author} {\bibfnamefont {T.}~\bibnamefont {Sawada}}, \
  and\ \bibinfo {author} {\bibfnamefont {M.}~\bibnamefont {Huang}},\ }\href
  {\doibase 10.1016/j.physletb.2020.135470} {\bibfield  {journal} {\bibinfo
  {journal} {Phys. Lett. B}\ }\textbf {\bibinfo {volume} {805}},\ \bibinfo
  {pages} {135470} (\bibinfo {year} {2020})},\ \Eprint
  {http://arxiv.org/abs/1910.10008} {arXiv:1910.10008 [hep-ph]} \BibitemShut
  {NoStop}%
\bibitem [{\citenamefont {Xie}\ \emph {et~al.}(2019)\citenamefont {Xie},
  \citenamefont {Watanabe},\ and\ \citenamefont {Huang}}]{Xie:2019soz}%
  \BibitemOpen
  \bibfield  {author} {\bibinfo {author} {\bibfnamefont {W.}~\bibnamefont
  {Xie}}, \bibinfo {author} {\bibfnamefont {A.}~\bibnamefont {Watanabe}}, \
  and\ \bibinfo {author} {\bibfnamefont {M.}~\bibnamefont {Huang}},\ }\href
  {\doibase 10.1007/JHEP10(2019)053} {\bibfield  {journal} {\bibinfo  {journal}
  {JHEP}\ }\textbf {\bibinfo {volume} {10}},\ \bibinfo {pages} {053} (\bibinfo
  {year} {2019})},\ \Eprint {http://arxiv.org/abs/1901.09564} {arXiv:1901.09564
  [hep-ph]} \BibitemShut {NoStop}%
\bibitem [{\citenamefont {Liu}\ \emph {et~al.}(2022)\citenamefont {Liu},
  \citenamefont {Xie}, \citenamefont {Sun}, \citenamefont {Li},\ and\
  \citenamefont {Watanabe}}]{Liu:2022out}%
  \BibitemOpen
  \bibfield  {author} {\bibinfo {author} {\bibfnamefont {Z.}~\bibnamefont
  {Liu}}, \bibinfo {author} {\bibfnamefont {W.}~\bibnamefont {Xie}}, \bibinfo
  {author} {\bibfnamefont {F.}~\bibnamefont {Sun}}, \bibinfo {author}
  {\bibfnamefont {S.}~\bibnamefont {Li}}, \ and\ \bibinfo {author}
  {\bibfnamefont {A.}~\bibnamefont {Watanabe}},\ }\href {\doibase
  10.1103/PhysRevD.106.054025} {\bibfield  {journal} {\bibinfo  {journal}
  {Phys. Rev. D}\ }\textbf {\bibinfo {volume} {106}},\ \bibinfo {pages}
  {054025} (\bibinfo {year} {2022})},\ \Eprint
  {http://arxiv.org/abs/2202.08013} {arXiv:2202.08013 [hep-ph]} \BibitemShut
  {NoStop}%
\bibitem [{\citenamefont {Liu}\ \emph {et~al.}(2023)\citenamefont {Liu},
  \citenamefont {Xie},\ and\ \citenamefont {Watanabe}}]{Liu:2022zsa}%
  \BibitemOpen
  \bibfield  {author} {\bibinfo {author} {\bibfnamefont {Z.}~\bibnamefont
  {Liu}}, \bibinfo {author} {\bibfnamefont {W.}~\bibnamefont {Xie}}, \ and\
  \bibinfo {author} {\bibfnamefont {A.}~\bibnamefont {Watanabe}},\ }\href
  {\doibase 10.1103/PhysRevD.107.014018} {\bibfield  {journal} {\bibinfo
  {journal} {Phys. Rev. D}\ }\textbf {\bibinfo {volume} {107}},\ \bibinfo
  {pages} {014018} (\bibinfo {year} {2023})},\ \Eprint
  {http://arxiv.org/abs/2210.11246} {arXiv:2210.11246 [hep-ph]} \BibitemShut
  {NoStop}%
\bibitem [{\citenamefont {Liu}\ and\ \citenamefont
  {Watanabe}(2023)}]{Liu:2023tjr}%
  \BibitemOpen
  \bibfield  {author} {\bibinfo {author} {\bibfnamefont {Z.}~\bibnamefont
  {Liu}}\ and\ \bibinfo {author} {\bibfnamefont {A.}~\bibnamefont {Watanabe}},\
  }\href {\doibase 10.1103/PhysRevD.108.034010} {\bibfield  {journal} {\bibinfo
   {journal} {Phys. Rev. D}\ }\textbf {\bibinfo {volume} {108}},\ \bibinfo
  {pages} {034010} (\bibinfo {year} {2023})},\ \Eprint
  {http://arxiv.org/abs/2306.00564} {arXiv:2306.00564 [hep-ph]} \BibitemShut
  {NoStop}%
\bibitem [{\citenamefont {Watanabe}\ \emph {et~al.}(2023)\citenamefont
  {Watanabe}, \citenamefont {Sirat},\ and\ \citenamefont
  {Liu}}]{Watanabe:2023rgp}%
  \BibitemOpen
  \bibfield  {author} {\bibinfo {author} {\bibfnamefont {A.}~\bibnamefont
  {Watanabe}}, \bibinfo {author} {\bibfnamefont {S.~A.}\ \bibnamefont {Sirat}},
  \ and\ \bibinfo {author} {\bibfnamefont {Z.}~\bibnamefont {Liu}},\ }\href
  {\doibase 10.1140/epjc/s10052-023-12062-0} {\bibfield  {journal} {\bibinfo
  {journal} {Eur. Phys. J. C}\ }\textbf {\bibinfo {volume} {83}},\ \bibinfo
  {pages} {898} (\bibinfo {year} {2023})},\ \Eprint
  {http://arxiv.org/abs/2305.06700} {arXiv:2305.06700 [hep-ph]} \BibitemShut
  {NoStop}%
\bibitem [{\citenamefont {Zhang}\ \emph {et~al.}(2023)\citenamefont {Zhang},
  \citenamefont {Chen}, \citenamefont {Li},\ and\ \citenamefont
  {Watanabe}}]{Zhang:2023nsk}%
  \BibitemOpen
  \bibfield  {author} {\bibinfo {author} {\bibfnamefont {Y.-P.}\ \bibnamefont
  {Zhang}}, \bibinfo {author} {\bibfnamefont {X.}~\bibnamefont {Chen}},
  \bibinfo {author} {\bibfnamefont {X.-H.}\ \bibnamefont {Li}}, \ and\ \bibinfo
  {author} {\bibfnamefont {A.}~\bibnamefont {Watanabe}},\ }\href {\doibase
  10.1103/PhysRevD.108.066001} {\bibfield  {journal} {\bibinfo  {journal}
  {Phys. Rev. D}\ }\textbf {\bibinfo {volume} {108}},\ \bibinfo {pages}
  {066001} (\bibinfo {year} {2023})},\ \Eprint
  {http://arxiv.org/abs/2307.00745} {arXiv:2307.00745 [hep-ph]} \BibitemShut
  {NoStop}%
\bibitem [{\citenamefont {Watanabe}\ \emph {et~al.}(2024)\citenamefont
  {Watanabe}, \citenamefont {Ahmadi}, \citenamefont {Liu},\ and\ \citenamefont
  {Xie}}]{Watanabe:2023kki}%
  \BibitemOpen
  \bibfield  {author} {\bibinfo {author} {\bibfnamefont {A.}~\bibnamefont
  {Watanabe}}, \bibinfo {author} {\bibfnamefont {Z.}~\bibnamefont {Ahmadi}},
  \bibinfo {author} {\bibfnamefont {Z.}~\bibnamefont {Liu}}, \ and\ \bibinfo
  {author} {\bibfnamefont {W.}~\bibnamefont {Xie}},\ }\href {\doibase
  10.1016/j.physletb.2023.138343} {\bibfield  {journal} {\bibinfo  {journal}
  {Phys. Lett. B}\ }\textbf {\bibinfo {volume} {848}},\ \bibinfo {pages}
  {138343} (\bibinfo {year} {2024})},\ \Eprint
  {http://arxiv.org/abs/2306.12890} {arXiv:2306.12890 [hep-ph]} \BibitemShut
  {NoStop}%
\bibitem [{\citenamefont {Zhang}\ \emph {et~al.}(2024)\citenamefont {Zhang},
  \citenamefont {Chen}, \citenamefont {Li},\ and\ \citenamefont
  {Watanabe}}]{Zhang:2024psj}%
  \BibitemOpen
  \bibfield  {author} {\bibinfo {author} {\bibfnamefont {Y.-P.}\ \bibnamefont
  {Zhang}}, \bibinfo {author} {\bibfnamefont {X.}~\bibnamefont {Chen}},
  \bibinfo {author} {\bibfnamefont {X.-H.}\ \bibnamefont {Li}}, \ and\ \bibinfo
  {author} {\bibfnamefont {A.}~\bibnamefont {Watanabe}},\ }\href {\doibase
  10.1103/PhysRevD.109.074010} {\bibfield  {journal} {\bibinfo  {journal}
  {Phys. Rev. D}\ }\textbf {\bibinfo {volume} {109}},\ \bibinfo {pages}
  {074010} (\bibinfo {year} {2024})},\ \Eprint
  {http://arxiv.org/abs/2401.14649} {arXiv:2401.14649 [hep-ph]} \BibitemShut
  {NoStop}%
\bibitem [{\citenamefont {Aguilar}\ \emph {et~al.}(2019)\citenamefont {Aguilar}
  \emph {et~al.}}]{Aguilar:2019teb}%
  \BibitemOpen
  \bibfield  {author} {\bibinfo {author} {\bibfnamefont {A.~C.}\ \bibnamefont
  {Aguilar}} \emph {et~al.},\ }\href {\doibase 10.1140/epja/i2019-12885-0}
  {\bibfield  {journal} {\bibinfo  {journal} {Eur. Phys. J. A}\ }\textbf
  {\bibinfo {volume} {55}},\ \bibinfo {pages} {190} (\bibinfo {year} {2019})},\
  \Eprint {http://arxiv.org/abs/1907.08218} {arXiv:1907.08218 [nucl-ex]}
  \BibitemShut {NoStop}%
\bibitem [{\citenamefont {Roberts}\ \emph {et~al.}(2021)\citenamefont
  {Roberts}, \citenamefont {Richards}, \citenamefont {Horn},\ and\
  \citenamefont {Chang}}]{Roberts:2021nhw}%
  \BibitemOpen
  \bibfield  {author} {\bibinfo {author} {\bibfnamefont {C.~D.}\ \bibnamefont
  {Roberts}}, \bibinfo {author} {\bibfnamefont {D.~G.}\ \bibnamefont
  {Richards}}, \bibinfo {author} {\bibfnamefont {T.}~\bibnamefont {Horn}}, \
  and\ \bibinfo {author} {\bibfnamefont {L.}~\bibnamefont {Chang}},\ }\href
  {\doibase 10.1016/j.ppnp.2021.103883} {\bibfield  {journal} {\bibinfo
  {journal} {Prog. Part. Nucl. Phys.}\ }\textbf {\bibinfo {volume} {120}},\
  \bibinfo {pages} {103883} (\bibinfo {year} {2021})},\ \Eprint
  {http://arxiv.org/abs/2102.01765} {arXiv:2102.01765 [hep-ph]} \BibitemShut
  {NoStop}%
\bibitem [{\citenamefont {Accardi}\ \emph {et~al.}(2016)\citenamefont {Accardi}
  \emph {et~al.}}]{Accardi:2012qut}%
  \BibitemOpen
  \bibfield  {author} {\bibinfo {author} {\bibfnamefont {A.}~\bibnamefont
  {Accardi}} \emph {et~al.},\ }\href {\doibase 10.1140/epja/i2016-16268-9}
  {\bibfield  {journal} {\bibinfo  {journal} {Eur. Phys. J. A}\ }\textbf
  {\bibinfo {volume} {52}},\ \bibinfo {pages} {268} (\bibinfo {year} {2016})},\
  \Eprint {http://arxiv.org/abs/1212.1701} {arXiv:1212.1701 [nucl-ex]}
  \BibitemShut {NoStop}%
\bibitem [{\citenamefont {Aschenauer}\ \emph {et~al.}(2019)\citenamefont
  {Aschenauer}, \citenamefont {Fazio}, \citenamefont {Lee}, \citenamefont
  {Mantysaari}, \citenamefont {Page}, \citenamefont {Schenke}, \citenamefont
  {Ullrich}, \citenamefont {Venugopalan},\ and\ \citenamefont
  {Zurita}}]{Aschenauer:2017jsk}%
  \BibitemOpen
  \bibfield  {author} {\bibinfo {author} {\bibfnamefont {E.~C.}\ \bibnamefont
  {Aschenauer}}, \bibinfo {author} {\bibfnamefont {S.}~\bibnamefont {Fazio}},
  \bibinfo {author} {\bibfnamefont {J.~H.}\ \bibnamefont {Lee}}, \bibinfo
  {author} {\bibfnamefont {H.}~\bibnamefont {Mantysaari}}, \bibinfo {author}
  {\bibfnamefont {B.~S.}\ \bibnamefont {Page}}, \bibinfo {author}
  {\bibfnamefont {B.}~\bibnamefont {Schenke}}, \bibinfo {author} {\bibfnamefont
  {T.}~\bibnamefont {Ullrich}}, \bibinfo {author} {\bibfnamefont
  {R.}~\bibnamefont {Venugopalan}}, \ and\ \bibinfo {author} {\bibfnamefont
  {P.}~\bibnamefont {Zurita}},\ }\href {\doibase 10.1088/1361-6633/aaf216}
  {\bibfield  {journal} {\bibinfo  {journal} {Rept. Prog. Phys.}\ }\textbf
  {\bibinfo {volume} {82}},\ \bibinfo {pages} {024301} (\bibinfo {year}
  {2019})},\ \Eprint {http://arxiv.org/abs/1708.01527} {arXiv:1708.01527
  [nucl-ex]} \BibitemShut {NoStop}%
\bibitem [{\citenamefont {Chen}\ \emph {et~al.}(2020)\citenamefont {Chen},
  \citenamefont {Guo}, \citenamefont {Roberts},\ and\ \citenamefont
  {Wang}}]{Chen:2020ijn}%
  \BibitemOpen
  \bibfield  {author} {\bibinfo {author} {\bibfnamefont {X.}~\bibnamefont
  {Chen}}, \bibinfo {author} {\bibfnamefont {F.-K.}\ \bibnamefont {Guo}},
  \bibinfo {author} {\bibfnamefont {C.~D.}\ \bibnamefont {Roberts}}, \ and\
  \bibinfo {author} {\bibfnamefont {R.}~\bibnamefont {Wang}},\ }\href {\doibase
  10.1007/s00601-020-01574-0} {\bibfield  {journal} {\bibinfo  {journal} {Few
  Body Syst.}\ }\textbf {\bibinfo {volume} {61}},\ \bibinfo {pages} {43}
  (\bibinfo {year} {2020})},\ \Eprint {http://arxiv.org/abs/2008.00102}
  {arXiv:2008.00102 [hep-ph]} \BibitemShut {NoStop}%
\bibitem [{\citenamefont {Abdul~Khalek}\ \emph {et~al.}(2022)\citenamefont
  {Abdul~Khalek} \emph {et~al.}}]{AbdulKhalek:2021gbh}%
  \BibitemOpen
  \bibfield  {author} {\bibinfo {author} {\bibfnamefont {R.}~\bibnamefont
  {Abdul~Khalek}} \emph {et~al.},\ }\href {\doibase
  10.1016/j.nuclphysa.2022.122447} {\bibfield  {journal} {\bibinfo  {journal}
  {Nucl. Phys. A}\ }\textbf {\bibinfo {volume} {1026}},\ \bibinfo {pages}
  {122447} (\bibinfo {year} {2022})},\ \Eprint
  {http://arxiv.org/abs/2103.05419} {arXiv:2103.05419 [physics.ins-det]}
  \BibitemShut {NoStop}%
\bibitem [{\citenamefont {Anderle}\ \emph {et~al.}(2021)\citenamefont {Anderle}
  \emph {et~al.}}]{Anderle:2021wcy}%
  \BibitemOpen
  \bibfield  {author} {\bibinfo {author} {\bibfnamefont {D.~P.}\ \bibnamefont
  {Anderle}} \emph {et~al.},\ }\href {\doibase 10.1007/s11467-021-1062-0}
  {\bibfield  {journal} {\bibinfo  {journal} {Front. Phys. (Beijing)}\ }\textbf
  {\bibinfo {volume} {16}},\ \bibinfo {pages} {64701} (\bibinfo {year}
  {2021})},\ \Eprint {http://arxiv.org/abs/2102.09222} {arXiv:2102.09222
  [nucl-ex]} \BibitemShut {NoStop}%
\bibitem [{\citenamefont {Lepage}\ and\ \citenamefont
  {Brodsky}(1980)}]{Lepage:1980fj}%
  \BibitemOpen
  \bibfield  {author} {\bibinfo {author} {\bibfnamefont {G.~P.}\ \bibnamefont
  {Lepage}}\ and\ \bibinfo {author} {\bibfnamefont {S.~J.}\ \bibnamefont
  {Brodsky}},\ }\href {\doibase 10.1103/PhysRevD.22.2157} {\bibfield  {journal}
  {\bibinfo  {journal} {Phys. Rev. D}\ }\textbf {\bibinfo {volume} {22}},\
  \bibinfo {pages} {2157} (\bibinfo {year} {1980})}\BibitemShut {NoStop}%
\bibitem [{\citenamefont {Tanaka}(2018)}]{Tanaka:2018wea}%
  \BibitemOpen
  \bibfield  {author} {\bibinfo {author} {\bibfnamefont {K.}~\bibnamefont
  {Tanaka}},\ }\href {\doibase 10.1103/PhysRevD.98.034009} {\bibfield
  {journal} {\bibinfo  {journal} {Phys. Rev. D}\ }\textbf {\bibinfo {volume}
  {98}},\ \bibinfo {pages} {034009} (\bibinfo {year} {2018})},\ \Eprint
  {http://arxiv.org/abs/1806.10591} {arXiv:1806.10591 [hep-ph]} \BibitemShut
  {NoStop}%
\bibitem [{\citenamefont {Tong}\ \emph {et~al.}(2022)\citenamefont {Tong},
  \citenamefont {Ma},\ and\ \citenamefont {Yuan}}]{Tong:2022zax}%
  \BibitemOpen
  \bibfield  {author} {\bibinfo {author} {\bibfnamefont {X.-B.}\ \bibnamefont
  {Tong}}, \bibinfo {author} {\bibfnamefont {J.-P.}\ \bibnamefont {Ma}}, \ and\
  \bibinfo {author} {\bibfnamefont {F.}~\bibnamefont {Yuan}},\ }\href {\doibase
  10.1007/JHEP10(2022)046} {\bibfield  {journal} {\bibinfo  {journal} {JHEP}\
  }\textbf {\bibinfo {volume} {10}},\ \bibinfo {pages} {046} (\bibinfo {year}
  {2022})},\ \Eprint {http://arxiv.org/abs/2203.13493} {arXiv:2203.13493
  [hep-ph]} \BibitemShut {NoStop}%
\bibitem [{\citenamefont {Workman}\ \emph {et~al.}(2022)\citenamefont {Workman}
  \emph {et~al.}}]{ParticleDataGroup:2022pth}%
  \BibitemOpen
  \bibfield  {author} {\bibinfo {author} {\bibfnamefont {R.~L.}\ \bibnamefont
  {Workman}} \emph {et~al.} (\bibinfo {collaboration} {Particle Data Group}),\
  }\href {\doibase 10.1093/ptep/ptac097} {\bibfield  {journal} {\bibinfo
  {journal} {PTEP}\ }\textbf {\bibinfo {volume} {2022}},\ \bibinfo {pages}
  {083C01} (\bibinfo {year} {2022})}\BibitemShut {NoStop}%
\end{thebibliography}%

\end{document}